\documentclass[reqno,12pt]{article}
\usepackage{amsmath,amsfonts,amssymb,amsthm,amstext,amscd,eucal,xcolor}
\usepackage[all]{xy}
\usepackage{cite,hyperref}
\usepackage{graphicx}
\usepackage[active]{srcltx}
\makeatletter \@addtoreset{equation}{section}

\makeatletter\renewcommand\section{\@startsection {section}{1}{\z@}%
                                   {-3.5ex \@plus -1ex \@minus -.2ex}%nn
                                   {2.3ex \@plus.2ex}%
                                   {\normalfont\large\bfseries}}
\renewcommand\subsection{\@startsection{subsection}{2}{\z@}%
                                     {-3.25ex\@plus -1ex \@minus -.2ex}%
                                     {1.5ex \@plus .2ex}%
                                     {\normalfont\bfseries}}

\newcommand{\be}{\begin{equation}}
\newcommand{\ee}{\end{equation}}
\newcommand{\bea}{\begin{eqnarray}}
\newcommand{\eea}{\end{eqnarray}}
\newcommand{\bse}{\begin{subequations}}
\newcommand{\ese}{\end{subequations}}
\newcommand{\beqa}{\begin{eqnarray}}
\newcommand{\eeqa}{\end{eqnarray}}
\newcommand{\beqar}{\begin{eqnarray*}}
\newcommand{\eeqar}{\end{eqnarray*}}
\newcommand{\bi}{\begin{itemize}}
\newcommand{\ei}{\end{itemize}}
\newcommand{\bn}{\begin{enumerate}}
\newcommand{\en}{\end{enumerate}}
\newcommand{\ba}{\begin{array}}
\newcommand{\ea}{\end{array}}
\newcommand{\bc}{\begin{center}}
\newcommand{\ec}{\end{center}}
\newcommand{\nnr}{\nonumber \\}

\definecolor{darkgreen}{rgb}{0,0.3,0}
\definecolor{darkblue}{rgb}{0,0,0.3}
\definecolor{darkred}{rgb}{0.7,0,0}

\begin{document}

\newcommand{\NC}{noncommutative }
\newcommand{\email}[1]{\footnote{E-mail: \href{mailto:#1}{#1}}}

\title{\textbf{\Large
{Three-dimensional noncommutative Yukawa theory:\\
Induced effective action and propagating modes}}}

\author{\textbf{R.~Bufalo}$^{1}$\email{rodrigo.bufalo@dfi.ufla.br} \textbf{and M. Ghasemkhani}$^{2}$\email{ghasemkhani@ipm.ir} \\
%EndAName
\textit{$^{1}$ \small Departamento de F\'isica, Universidade Federal de Lavras,}\\
\textit{ \small Caixa Postal 3037, 37200-000 Lavras, MG, Brazil}\\
\textit{$^{2}$ \small Department of Physics, Shahid Beheshti University,}\\
\textit{ \small G.C., Evin, Tehran 19839, Iran}\\
}

 \maketitle
\begin{abstract}
In this paper we establish the analysis of \NC Yukawa theory, encompassing neutral and charged scalar fields. We approach the analysis by considering
carefully the derivation of the respective effective actions. Hence,
based on the obtained results, we compute the one-loop contributions
to the neutral and charged scalar field self-energy, as well as to the
Chern-Simons polarization tensor. In order to properly define
the behaviour of the quantum fields, the known UV/IR mixing due to radiative corrections is
analysed in the one-loop physical dispersion relation of the scalar
and gauge fields.
\end{abstract}

\begin{flushleft}
{\bf PACS:} 11.15.-q, 11.10.Kk, 11.10.Nx
\end{flushleft}

%%%%%%%%%%%%%%%%%%%%%%%%%%%%%%%%%%%%%%%%%%%%%%%%%%%%%%%%%%%%%%%%%%%%%%%%%%%%%%%%%%%%%%%%%%%%%%%%%%%%%%%%%%%%%%%%%%%%%%%%%%%%%%%%%%%%%%%%%%%%%%%%%%%%%%%%%%%
\newpage
\tableofcontents
\newpage
%%%%%%%%%%%%%%%%%%%%%%%%%%%%%%%%%%%%%%%%%%%%%%%%%%%%%%%%%%%%%%%%%%%%%%%%%%%%%%%%%%%%%%%%%%
\section{Introduction}
\label{sec1}
%%%%%%%%%%%%%%%%%%%%%%%%%%%%%%%%%%%%%%%%%%%%%%%%%%%%%%%%%%%%%%%%%%%%%%%%%%%%%%%%%%%%%%%%
The field theoretical model for description of the interaction between nucleons in particle physics was first proposed by H.~Yukawa in 1935 \cite{yukawa}, which led to the prediction of pion before its discovery from cosmic rays in 1947 \cite{powell}. The Yukawa term originates from the exchange of a massive scalar field that in the non-relativistic limit yields a Yukawa potential and hence the corresponding force has a finite range, which is inversely proportional to the mediator particle mass.

Since its proposal, the notion of Yukawa potential has been used in different areas in the description of several phenomena such as chemical process, astrophysics, fluid plasma system and especially in modern particle physics. More importantly, in the latter case, i.e. standard model, the Yukawa interaction of the Higgs field and massless quarks and leptons is the responsible coupling to give mass to these fermionic fields.

Due to its importance in the different physical phenomena, Yukawa theory has been used as a laboratory  in the search of physics beyond standard model, or even to scrutinize the cornerstones of gauge theories. 
Furthermore, if we expand our
scope and add to our interest the description of nature
behavior at shortest distances \cite{ref23,ref24}, i.e. a quantum
theory of gravity, or even the so-called minimal length scale
physics, one inexorably finds that \NC geometry is one of the most
highly motivated and richer framework \cite{ref16}, including
phenomenological inspirations
\cite{Chaichian:2001py,ref17,Nicolini:2008aj,ref23,Amelino-Camelia:2013fxa}.
Space-time noncommutativity naturally emerges at Plank scale in
attempts to accommodate quantum mechanics and general relativity in
a common framework, one finds uncertainty principles that are
compatible with non-commuting coordinates
\cite{doplicher,Ahluwalia:1993dd}.

The simplest realization of
noncommutativity is given by the following canonical algebra
\begin{equation}
\left[\hat{x}_\mu , \hat{x}_\nu\right] = i\theta_{\mu \nu}.
\label{eq:0.01}
\end{equation}
where $\theta_{\mu\nu}$ is a constant skewsymmetric matrix of
dimension of length squared. These commutation relations give an
uncertainty relation among the coordinates: $ \Delta\hat{x}_\mu
~\Delta\hat{x}_\nu \gtrsim \frac{1}{2} |\theta_{\mu \nu}|$. Notice,
however, that the nonzero components of $\theta^{\mu \nu}$ are
arbitrary parameters that must be constrained by experiments, we can
think as they being the resolution scale that can be taken to be,
for instance, of the order of square of Planck length $\ell _{Pl}$.

A suitable framework to compute quantities in a NC--QFT is by the
use of Weyl-Moyal (symbol) correspondence \cite{Wignerfunctions}.
This allows to define a classical (commutative) analogue of the \NC
space, so that the following relation holds: $\varphi
\left(\hat{x}\right) \psi \left(\hat{x}\right) \rightarrow \varphi
_W \star \psi_W $, where $\varphi _W $ is the so-called Weyl symbol
of the operator $\varphi \left(\hat{x}\right)$
\cite{Wignerfunctions}. Moreover, in this context, we have that the
Moyal star product is defined as
 \begin{equation}
f(x)\star
g(x)=f(x)\exp\left(\frac{i}{2}\theta^{\mu\nu}\overleftarrow{\partial}_{\mu}
\overrightarrow{\partial}_{\nu}\right)
g(x).
 \end{equation}

One common property of NC gauge theories that has been uncovered is
that high-momentum modes (UV) affect the physics at large distances
(IR) leading to the appearance of the so-called UV/IR mixing
\cite{minwalla}. These ``anomalies'' involve non-analytic behavior
in the noncommutativity parameter $\theta$ making the limit $\theta
\rightarrow0$ singular. Despite of the many attempts to understand
this issue in four and three-dimensional field theory models, see
\cite{D'Ascanio:2016asa} and
\cite{Vitale:2012dz,Gere:2015ota}, respectively, no complete
description to handle it has yet been provided
\cite{Blaschke:2009rb}.

An important branch of interest regarding NC gauge theories is the
study of how noncommutativity affects established properties of
conventional theories, in particular a considerable effort has been
expended in analyzing gauge theories defined in a three-dimensional
noncommutative spacetime, this effort is highly supported by the fact that
wandering into lower-dimensional models has been proved to be very
fertile and stimulated significantly the development of our
knowledge in the subject. Gauge theories defined in a
three-dimensional spacetime \cite{jackiw} are
known to possess unique properties and are well motivated as
providing a simple setting where important theoretical ideas are
suitably tested. Noncommutative three-dimensional field theory, in particular
gauge theory, can find application in the study of planar physics in
condensed matter and statistical physics
\cite{susskind,Bastos:2012kh,Sidharth:2014xya}.
After this observation, various perturbative aspects of the \NC
Chern-Simons theory have been studied
\cite{chen,bichl,das,jabbari-0,banerjee,Asano}, NC
Maxwell-Chern-Simons theory \cite{ref21,ghasemkhani} and NC
$QED_{3}$ \cite{ref13,bufalo}, as well as its supersymmetric
extension\cite{Ferrari}, where deviations of known phenomena and
interesting new properties have been uncovered.

However, as we have extensively discussed, in addition to its importance in the Standard Model of particles,
a \NC extension of the Yukawa field theory action should be fully
considered, in particular how the \NC Higgs effective action can be
generated. Some aspects for this theory have been discussed
previously
\cite{Panigrahi:2004cf,Blaschke:2005dv,Martin:2006gw,Horvat:2011qn,Bouchachia:2015kxa}.

In particular, our present analysis will be twofold: first, we will
consider the interaction between a neutral scalar field and
dynamical fermionic fields, where the scalar field effective action
is found by integrating out the fermionic modes, an additional
derivative cubic coupling is found for the scalar field. Second, a
more interesting case is considered, now we have the interaction
among charged scalar fields with fermionic fields augmented by gauge
fields, in which the effective action describing the interaction
between the charged scalar and gauge fields is obtained. In the
latter, we shall consider the dynamics of the gauge sector given by
the higher-derivative (HD) Chern-Simons action \cite{bufalo}, where
new features are discussed. At last, in both cases, UV/IR mixing is
analysed in the one-loop physical dispersion relation of the scalar
and gauge fields due to radiative corrections,
this is justifiable once this anomaly might modify significantly the
behavior of the quantum field in the description of a given
phenomen and find room in interesting
application \cite{Martin-Ruiz:2015skg}.

Therefore, in this paper we will consider the effective action for
two distinct Yukawa couplings: i) neutral scalar field with
fermionic fields, and ii) charged scalar field with fermionic fields
plus a gauge field. For this purpose, we will make use of the ideas
outlined in Ref.~\cite{ref11,bufalo}, in which noncommutative
fermionic effective actions were considered. The cornerstone of this
approach is outlined in Sec.~\ref{sec2} and consists in consider, in
a formal way, the existence of an exact Seiberg--Witten map
\cite{witten}, valid to all orders in $\theta$ so that the
noncommutative effects into the resulting outcome are in fact
nonperturbative. We compute explicitly in section \ref{sec3} the
respective effective action for the neutral and charged scalar
field, where the presence of the new couplings is discussed. Based on
the obtained results for the effective action, we proceed in Sec.~\ref{sec4} to compute the one-loop correction to the self-energy of the neutral scalar field, in particular application to its dispersion relation.
Next, in Sec.~\ref{sec5}, we consider the effective action obtained
for the case of charged scalar fields minimally coupled with a HD
Chern-Simons gauge field, a model that one can name as NC
HD-Chern-Simons-Higgs model. In this case we carefully analyse the
dispersion relation for both fields, by computing the set of
diagrams for the respective one-loop self energy functions.
 Additionally to what we have already discussed, another physical application where the present study can be employed, follows from Refs.~\cite{bak,boz1,boz2} in which it is shown that a model where the Chern-Simons term coupled to the scalar field matter is a suitable framework for field theoretical description of the Aharonov-Bohm effect. At
Sec.~\ref{sec6} we summarize our results and present our conclusion
and prospects.

%%%%%%%%%%%%%%%%%%%%%%%%%%%%%%%%%%%%%%%%%%%%%%%%%%%%%%%%%
\section{General discussion}
\label{sec2}
%%%%%%%%%%%%%%%%%%%%%%%%%%%%%%%%%%%%%%%%%%%%%%%%%%%%%%%%%%
Let us now define the noncommutative extension of fermionic fields
interacting with a neutral scalar field, which can be named as a
noncommutative Yukawa model. For this, we shall consider the
following action
\begin{equation}
S=\int d^{3}x\left[i\bar{\psi}\star
\gamma^{\mu}\partial_{\mu}\psi-m\bar{\psi}\star\psi+g\bar{\psi}\star\phi\star\psi\right].\label{eq:0.1}
\end{equation}
It should be remarked that we are working with a two-component
representation for the spinors with the standard convention
\begin{equation}
\gamma^{0}=\sigma^{3}=\left(
                        \begin{array}{cc}
                          1 & 0\\
                          0 & -1 \\
                        \end{array}
                      \right),\quad\quad
\gamma^{1}=i\sigma^{1}= \left(
                        \begin{array}{cc}
                          0 & i\\
                          i & 0 \\
                        \end{array}
                       \right),\quad\quad
 \gamma^{2}=i\sigma^{2}= \left(
                        \begin{array}{cc}
                          0 & 1\\
                          -1 & 0 \\
                        \end{array}
                       \right),\label{eq:0.02}
\end{equation}
where the $\gamma$-matrices satisfy
$\gamma^{\mu}\gamma^{\nu}=\eta^{\mu\nu}-i\varepsilon^{\mu\nu\sigma}\gamma_{\sigma}$.
 We observe that the action \eqref{eq:0.1} is invariant under a global $U(1)$ symmetry, $\psi\rightarrow e^{i\alpha}\psi$. Furthermore, on the behavior of this action under discrete transformations, parity ($\textbf{P}$), charge conjugation ($\textbf{C}$) and time reversal ($\textbf{T}$), we have prepared a detailed analysis in the following:
\begin{itemize}
  \item [(\emph{i})]~\emph{Parity}\newline
  \newline
  The description of the parity transformation in $2+1$ dimensions is given by $x_{1}\rightarrow -x_{1}$ and $x_{2}\rightarrow x_{2}$. Using the invariance of the kinetic part of the Dirac Lagrangian under parity, it is found that the fermionic field transforms as $\psi\rightarrow\gamma^{1}\psi$. Hence, it is easily concluded that parity is broken by the fermion mass term, since $\bar\psi\psi\rightarrow-\bar\psi\psi$.

From \eqref{eq:0.01}, it is deduced that the \NC parameter changes sign under parity $\theta \rightarrow - \theta$, and hence  we observe that the interaction term $\bar{\psi}\star\phi\star\psi$ transforms into the following
 \begin{equation}
  S_{\emph{int}}^{P}=
  -g\int d^{3}x~ \bar\psi\star(\psi\star\phi_{_{P}}),
\end{equation}
in which we have used the anti-commuting property of the fermionic fields. If we assume a pseudo scalar field $\phi_{_{P}}=-\phi$, similar to the three-dimensional Yukawa term in commutative space, then it is shown that the Yukawa coupling as considered here Eq.~\eqref{eq:0.1} is not parity invariant; however, it follows that we can construct a combination that is parity invariant
\begin{equation}
\widetilde{S}_{\emph{int}}=g\int d^{3}x~\bar{\psi}\star\{\phi,\psi\}_{\star},
\end{equation}
here $\{~,~\}=[~,~]_{+}$. On the other hand, for the other choice $\phi_{_{P}}=+\phi$, it is easily realized that
\begin{equation}
 \widetilde{S}_{\emph{int}}=g\int d^{3}x~\bar{\psi}\star[\phi,\psi]_{\star},
\end{equation}
is parity invariant.
  \item [(\emph{ii})]~\emph{Charge conjugation}\newline
\newline
  Under a charge conjugation transformation in three dimensions, the spinor field changes as $\psi\rightarrow \mathbf{C}\gamma^{0}\psi^{\star}$, so that the operator $\textbf{C}$ should satisfy the following relation
  \begin{equation}
  \mathbf{C}^{-1}\gamma^{\mu}\mathbf{C}=-(\gamma^{\mu})^{T}.
  \end{equation}
  Hence, by considering the above constraint and also the representation of the gamma matrices in \eqref{eq:0.02}, the appropriate choice for the charge conjugation operator in $2+1$ dimensions is given by $\mathbf{C}=\gamma^{2}$. Since $(\gamma^{2})^{\star}=\gamma^{2}$ and $(\gamma^{0})^{\star}=\gamma^{0}$ then $\psi^{\star}\rightarrow \gamma^{2}\gamma^{0}\psi$, so that consequently we can find the transformation of the fermion mass term as follows
  \begin{equation}
  \bar\psi_{c}\psi_{c}   =-\psi^{T}\gamma^{0}\psi^{\star}
  =\bar\psi\psi,
  \end{equation}
  where $(\gamma^{2})^{T}=-\gamma^{2}$ and the anticommuting property of the spinors had been used.
  We thus see that the fermion mass term is $\textbf{C}$-invariant. Finally, if we consider that the scalar field satisfies $\phi_{c}=\phi$, the noncommutative Yukawa
  interaction term transforms as
  \begin{equation}
    S_{\emph{int}}^{c}= g\int d^{3}x~\bar{\psi}\star(\psi\star\phi)\label{cc}
\end{equation}
We notice that there are two different choices to build a noncommutative Yukawa interaction term that these are related to each other by a charge conjugation transformation. This point was first mentioned in \cite{Shahin}, including the study of the discrete symmetries in noncommutative  $QED_{4}$. Furthermore, if we apply $\theta\rightarrow -\theta$ to \eqref{cc}, we find that the interaction term is $\textbf{C}$ invariant.
  \item [(\emph{iii})]~\emph{Time reversal}\newline
  \newline
  Time reversal operator acts on the fermionic field as $\psi\rightarrow\gamma^{2}\psi$. Thus the fermion mass term, similar to the parity transformation, is not invariant under $\textbf{T}$, since it behaves as $\bar\psi\psi\rightarrow -\bar\psi\psi$. For the interaction term, we have
 \begin{equation}
  S_{\emph{int}}^{T}= g\int d^{3}x~\bar{\psi}\star(\psi\star\phi)
\end{equation}
in which it is supposed that $\phi_{T}=-\phi$. Once again we observe that, similar to the charge conjugation transformation, adding the assumption $\theta\rightarrow -\theta$ gives us a $\textbf{T}$-invariant interaction term.
\end{itemize}
Finally, we are able to establish the one-loop effective action for the
scalar field $\phi$ by integrating over the fermionic fields,
\begin{equation}
i\Gamma\left[\phi\right]=\ln\frac{\det\left(i\displaystyle{\not}\partial-m+g\phi\star\right)}
{\det\left(i\displaystyle{\not}\partial-m\right)}=
-\sum_{n}\frac{1}{n}tr\bigg(\left(\displaystyle{\not}\partial+im\right)^{-1}
i\left(g\phi\star\right)\bigg)^{n},\label{eq:
0.2}
\end{equation}
where we identify the differential operator as for the fermionic
propagator,
\begin{equation}
\left(\displaystyle{\not}\partial+im\right)^{-1}\delta\left(x-y\right)=
\int\frac{d^{3}p}{\left(2\pi\right)^{3}}\frac{i\left(\displaystyle{\not}p
+m\right)}{p^{2}-m^{2}+i\varepsilon}e^{-ip.\left(x-y\right)}.
\end{equation}
Nevertheless, due to our interest, we can rewrite \eqref{eq: 0.2} in
a far more convenient form as
\begin{equation}
i\Gamma\left[\phi\right]=\sum_{n}\int d^{3}x_{1}\ldots\int
d^{3}x_{n}\left[\phi\left(x_{1}\right)\phi\left(x_{2}\right)\cdots\phi\left(x_{n}\right)\right]
\Gamma\left(x_{1},x_{2},\ldots,x_{n}\right),\label{eq:0.3}
\end{equation}
where perturbative calculation is readily obtained and
we have defined
\begin{align}
\Gamma\left(x_{1},x_{2},\ldots
x_{n}\right)&=-\frac{\left(-g\right)^{n}}{n}\int\prod_{i}\frac{d^{3}p_{i}}{\left(2\pi\right)^{3}}
\left(2\pi\right)^{3}\delta\bigg(\sum_{i}p_{i}\bigg)
\nonumber \\
&\times\exp\left(-i\sum_{i}p_{i}x_{i}\right)\exp\left(-\frac{i}{2}\sum_{i<j}p_{i}\times
p_{j}\right)\Xi\left(p_{1},p_{2},\ldots,p_{n-1}\right),\label{eq:
0.4}
\end{align}
in which we introduced the notation $p\times
q=\theta^{\mu\nu}p_{\mu}q_{\nu}$; by simplicity, the
one-loop contributions are defined in the form
\begin{equation}
\Xi\left(p_{1},\ldots,p_{n-1}\right)=\int\frac{d^{3}q}{\left(2\pi\right)^{3}}
\frac{tr\bigg[\left(\displaystyle{\not}q+\displaystyle{\not}p_{1}+m\right)\left(\displaystyle{\not}q
+m\right)\left(\displaystyle{\not}q-\displaystyle{\not}p_{2}+m\right)
\ldots\bigg(\displaystyle{\not}q-\sum\limits_{i=2}^{n-1}\displaystyle{\not}p_{i}+m\bigg)\bigg]}
{\left[\left(q+p_{1}\right)^{2}-m^{2}\right]
\left[q^{2}-m^{2}\right]\left[\left(q-p_{2}\right)^{2}-m^{2}\right]\ldots
\bigg[\bigg(q-\sum\limits_{i=2}^{n-1}p_{i}\bigg)^{2}-m^{2}\bigg]}.\label{eq:
0.5}
\end{equation}
In order to rewrite \eqref{eq:0.3} into the form \eqref{eq: 0.5} we
have made use of the general result
\begin{align}
\int dx~\mathcal{O}_{1}\left(x\right)\star\mathcal{O}_{2}\left(x\right)\ldots &\star\mathcal{O}_{n}
\left(x\right)=\int\prod_{i}d^{3}x_{i}\prod_{i}\frac{d^{3}p_{i}}{\left(2\pi\right)^{3}}~
\mathcal{O}_{1}\left(x_{1}\right)\mathcal{O}_{2}\left(x_{2}\right)\ldots\mathcal{O}_{n}\left(x_{n}\right)
 \nonumber \\
&\times
\exp\bigg(-i\sum_{i}p_{i}x_{i}\bigg)\exp\bigg(-\frac{i}{2}\sum_{i<j}p_{i}\times
p_{j}\bigg)\delta\bigg(\sum_{i}p_{i}\bigg).\label{eq:
0.6}
\end{align}
With this result we finish our formal development where all the
necessary information were carefully presented. In the next section
we will proceed in computing explicitly the full effective action
for two cases: first for a neutral scalar field, and second for a
charged scalar field. In order to compute such contributions, we
will concentrate in considering the leading contributions for the
resulting expressions, this can be suitably achieved by means of the
long wavelength limit (i.e., $m^{2}>p^{2}$, where $p$ is an external
momentum).
%%%%%%%%%%%%%%%%%%%%%%%%%%%%%%%%%%%%%%%%%%%%%%%
\section{Perturbative effective action}
\label{sec3}
%%%%%%%%%%%%%%%%%%%%%%%%%%%%%%%%%%%%%%%%%%%%%%%%
\subsection{Neutral scalar fields}
%%%%%%%%%%%%%
We shall now proceed in evaluating explicitly the contributions of
two, three and four scalar fields to the effective action. Actually,
the contribution of one scalar field is identically
vanishing. In this case we will find that at the long wavelength
limit we generate a full action for the neutral scalar field, in
particular that no self-coupling is present at order higher than
three, only derivative couplings are available.
%%%%%%%
\subsubsection{\textmd{$\phi\phi$} contribution}
\begin{figure}[t]
\vspace{-0.5cm}
\includegraphics[width=4.8cm,height=2.5cm]{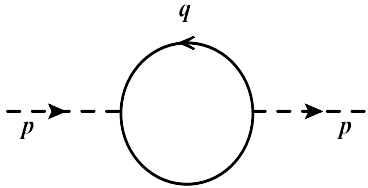}
 \centering\caption{Relevant graph for the induced $\phi\phi$-term.}
\label{oneloopdiagrams1}
\end{figure}

%%%%%%%
Let us consider the first nonvanishing contribution of the one-loop
effective action, for this matter we take $n=2$ in the Eq.\eqref{eq:
0.5}, depicted in Fig.~\ref{oneloopdiagrams1},
\begin{equation}
\Xi\left(p\right)=\int\frac{d^{3}q}{\left(2\pi\right)^{3}}\frac{tr\left[\left(\displaystyle{\not}q+\displaystyle{\not}p+m\right)
\left(\displaystyle{\not}q+m\right)\right]}{\left[\left(q+p\right)^{2}-m^{2}\right]\left[q^{2}-m^{2}\right]}.
\label{eq:0.9}
\end{equation}
%%%%%%%%%%%%%%%%%%%%%%%%%%%%%%%%%%%%%%%%%%%%%%%%%%%%%%
%%%%%%%%%%%%%%%%%%%%%%%%%%%%%%%%%%%%%%%%%%%%%%%%%%%%%%
%%%%%%%%%%%%%%%%%%%%%%%%%%%%%%%%%%%%%%%%%%%%%%%%%%%%%%
%%%%%%%%%%%%%%%%%%%%%%%%%%%%%%%%%%%%%%%%%%%%%%%%%%%%%%
Moreover, the momentum integration can be readily evaluated by
considering the Feynman parametrization of the denominator factors,
and considering the change of variables $q\rightarrow q+xp$, so that
\begin{equation}
\Xi\left(p\right)=\int_{0}^{1}dx\int\frac{d^{3}q}{\left(2\pi\right)^{3}}\frac{tr\left[\left(\displaystyle{\not}q+\left(1-x\right)
\displaystyle{\not}p+m\right)\left(\displaystyle{\not}q-x\displaystyle{\not}p+m\right)\right]}{\left[q^{2}+x\left(1-x\right)p^{2}-m^{2}\right]^{2}}.
\label{eq:0.10}
\end{equation}
The trace of $\gamma$-matrices in the numerator of \eqref{eq:0.10}
can be computed with help of the results
\begin{equation}
tr\left(\gamma^{\mu}\gamma^{\nu}\right)=2\eta^{\mu\nu},\quad
tr\left(\gamma^{\mu}\gamma^{\nu}\gamma^{\beta}\right)=2i\varepsilon^{\mu\nu\beta}.
\end{equation}
Finally, we write the above integral in a dimensional regularized
form as
\begin{equation}
\Xi\left(p\right)=2\int_{0}^{1}dx\int\frac{d^{\omega}q}{\left(2\pi\right)^{\omega}}
\frac{q^{2}-x\left(1-x\right)p^{2}+m^{2}}{\left[q^{2}+x\left(1-x\right)p^{2}-m^{2}\right]^{2}}.
\end{equation}
Hence, the integration in the momentum $q$ is rather straightforward
and the resulting expression has no poles when
$\omega\rightarrow3^{+}$, so the result reads
\begin{align}
\Xi\left(p\right) &
=\frac{i}{\pi}\int_{0}^{1}dx\sqrt{m^{2}-x\left(1-x\right)p^{2}}.\label{eq:
0.12}
\end{align}
We note that no scalar Chern--Simons-like term, proportional to $\varepsilon^{\mu\nu\beta}$, is generated.\footnote{ This result for a scalar field is in contrast to the effective action for a vector field, where an induced Chern--Simons action is produced in three-dimensional $QED$, at the large fermion mass limit \cite{bufalo}.} This is understood since it is impossible to build a Lorentz invariant quadratic combination of $\phi$ and $\varepsilon^{\mu\nu\beta}$ at this order.

\paragraph{Long wavelength limit}

Let us take a look at the remaining integration at the Eq.\eqref{eq:
0.12}. Moreover, considering the case when $p^{2}\ll m^{2}$, then we
find that
$\Xi\left(p\right)=-\frac{i}{12\pi}\frac{1}{\left|m\right|}\left(p^{2}-12m^{2}\right)$.
As it is easily seen, this $\mathcal{O}\left(m^{-1}\right)$ term
corresponds to the kinetic term of the Klein-Gordon action for the
neutral scalar field.

In the configuration space, if we replace the above result into the
expression \eqref{eq: 0.4} we find after some manipulation that
\begin{align}
\Gamma\left(x_{1},x_{2}\right) &
=-\frac{ig^{2}}{24\pi}\frac{1}{\left|m\right|}\left(\square+12m^{2}\right)\delta\left(x_{1}-x_{2}\right).
\end{align}
As it is well-known, there is no noncommutativity effects for the
case of two fields, since the phase factor in \eqref{eq: 0.4}
vanishes. Finally, the analysis of the first
nonvanishing term in \eqref{eq:0.3} is given by
\begin{align}
i\Gamma\left[\phi\phi\right] &
=\frac{ig^{2}}{24\pi}\frac{1}{\left|m\right|}\int
d^{3}x\left(\partial_{\mu}\phi\partial^{\mu}\phi-\mu^{2}\phi^2\right)\left(x\right),\label{eq:0.13}
\end{align}
and it leads to the radiatively induced Klein-Gordon action, where we have  introduced a new square mass parameter
$\mu^{2}=12m^{2}$.

%%%%%%%%%%%%
\subsubsection{$\phi\phi\phi$ contribution}
\begin{figure}[t]
%\vspace{-1.2cm}
\includegraphics[width=4.3cm,height=3.3cm]{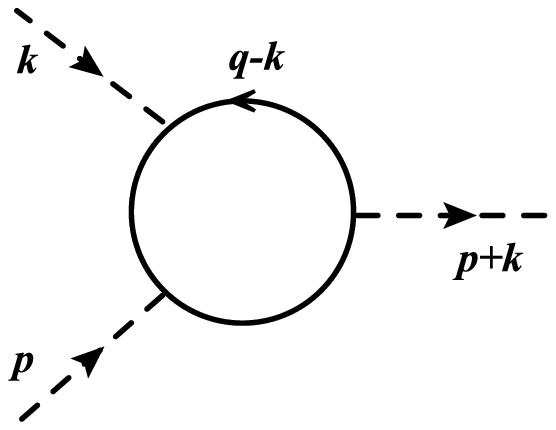}
 \centering  \caption{Relevant graph for the induced $\phi\phi\phi$-term.}
\label{oneloopdiagrams2}
\end{figure}

%%%%%%%%%%%%
The calculation of the next contribution follows as in the previous
analysis. We compute the $n=3$ contribution in the Eq.\eqref{eq:
0.5} and given in Fig.~\ref{oneloopdiagrams2},
\begin{equation}
\Xi\left(p,k\right)=\int\frac{d^{3}q}{\left(2\pi\right)^{3}}\frac{tr\left[\left(\displaystyle{\not}q+\displaystyle{\not}p+m\right)
\left(\displaystyle{\not}q+m\right)\left(\displaystyle{\not}q-\displaystyle{\not}k+m\right)\right]}{\left[\left(q+p\right)^{2}-m^{2}\right]\left[q^{2}-m^{2}\right]\left[\left(q-k\right)^{2}-m^{2}\right]}.\label{eq:
1.1}
\end{equation}
%Moreover, once again we can make use of the Feynman parametrization
%with the change of variables $q\rightarrow q-\left(xp-zk\right)=q-s$, so that
%\begin{equation}
%\Xi\left(p,k\right)=\Gamma\left(3\right)\int
%d\xi\int\frac{d^{\omega}q}{\left(2\pi\right)^{\omega}}
%\frac{tr\left[\left(\displaystyle{\not}q-\displaystyle{\not}s+\displaystyle{\not}p+m\right)
%\left(\displaystyle{\not}q-\displaystyle{\not}s+m\right)\left(\displaystyle{\not}q-\displaystyle{\not}s-
%\displaystyle{\not}k+m\right)\right]}{\left[q^{2}+A^{2}\left(p,k\right)-m^{2}\right]^{3}},\label{eq:
%1.3}
%\end{equation}
%%%%%%%%%%%%%%%%%%%%%%%%%%%%%%%%%%%%%%%%%%%%%%%%%%%%%%
%%%%%%%%%%%%%%%%%%%%%%%%%%%%%%%%%%%%%%%%%%%%%%%%%%%%%%
%%%%%%%%%%%%%%%%%%%%%%%%%%%%%%%%%%%%%%%%%%%%%%%%%%%%%%
%%%%%%%%%%%%%%%%%%%%%%%%%%%%%%%%%%%%%%%%%%%%%%%%%%%%%%
 The momentum integral can be computed straightforwardly
 using the dimensional regularization, and
realizing that $\Gamma\left(2-\frac{\omega}{2}\right)$ and
$\Gamma\left(3-\frac{\omega}{2}\right)$ have no poles when
$\omega\rightarrow3^{+}$, so we find
\begin{align}
\Xi\left(p,k\right) & =\frac{i}{16\pi}\int
d\xi\left[\frac{N_{1}\left(p,k;x,z\right)}{\left(m^{2}-A^{2}\left(p,k\right)\right)^{\frac{1}{2}}}
-\frac{N_{2}\left(p,k;x,z\right)}{\left(m^{2}-A^{2}\left(p,k\right)\right)^{\frac{3}{2}}}\right],\label{eq:
1.5}
\end{align}
where by simplicity we have defined
$s_{\mu}=xp_{\mu}-zk_{\mu}$ , the measure $\int
d\xi=\int_{0}^{1}dx\int_{0}^{1-x}dz$, and the quantity
$A^{2}\left(p,k\right)=-\left(xp-zk\right)^{2}+xp^{2}+zk^{2}$, as well as the following quantities
\begin{align}
N_{1}\left(p,k;x,z\right) &
=3tr\left(-3\displaystyle{\not}s-\displaystyle{\not}k+\displaystyle{\not}p\right)+18m,
\end{align}
and
\begin{align}
N_{2}\left(p,k;x,z\right) &
=-tr\left[\left(\displaystyle{\not}s-\displaystyle{\not}p\right)\left(\displaystyle{\not}s-m\right)\left(\displaystyle{\not}s+
\displaystyle{\not}k-m\right)\right]+mtr\left[\left(\displaystyle{\not}s-m\right)\left(\displaystyle{\not}s+\displaystyle{\not}k-m\right)\right].
\end{align}
As usual, the full contribution of \eqref{eq: 1.5} gives rise to a
full set of information, but here we are interested in the
particular cases of the $\mathcal{O}\left(m^{0}\right)$ and
$\mathcal{O}\left(m^{-2}\right)$ contributions, in order to add
these self-interacting terms to the kinetic contribution \eqref{eq:0.13}.

\paragraph{Derivative coupling}

Although the structure of the contribution \eqref{eq: 1.5} is rather
complicated than those from the two fields, we can keep traced of
the terms $\mathcal{O}\left(m^{0}\right)$ and
$\mathcal{O}\left(m^{-2}\right)$ by paying careful attention to the
contributions from the numerator and denominator at the long
wavelength limit. For that matter, we shall focus on the
$\mathcal{O}\left(m^{1}\right)$ and $\mathcal{O}\left(m^{3}\right)$
contributions from the quantities $N_{1}\left(p,k;x,z\right)$ and
$N_{2}\left(p,k;x,z\right)$,  and take the $p^{2}\ll m^{2}$ limit. With such considerations the remaining integral can now be computed,
so that the three fields contribution is simply given by
\begin{align}
\Xi\left(p,k\right) &
=\frac{i}{8\pi}\frac{m}{\left|m\right|}\left[4+\frac{1}{6m^{2}}\left(3k^{2}+2p^{2}+2\left(p.k\right)\right)\right].\label{eq:1.13}
\end{align}
Finally, the interacting effective action \eqref{eq:0.3} of the
noncommutative Klein-Gordon action for a neutral scalar field is
found to be
\begin{align}
i\Gamma\left[\phi\phi\phi\right] &
=\frac{ig^{3}}{6\pi}\frac{m}{\left|m\right|}\int
d^{3}x\bigg(\phi\star\phi\star\phi+\frac{1}{3m^{2}}
\partial^{\alpha}\phi\star\partial_{\alpha}\phi\star\phi\bigg)\left(x\right),\label{eq:
1.12}
\end{align}
where we have applied the identity $2\int d
^{3}x~\partial^{\alpha}\phi\star\partial_{\alpha}\phi\star\phi=-\int
d ^{3}x~\square\phi\star\phi\star\phi$, found as a result of using the
cyclic property of the Moyal product and performing an integration
by part. We thus see that a noncommutative
$\lambda\phi_{\star}^{3}$ interacting term is radiatively generated,
in addition to a derivative coupling as well, where the coupling
constant has dimension of $\left[\lambda\right]=\left[gm\right]$.
It is worth to mention that due to theory's structure higher
self-interacting contributions $\lambda\phi_{\star}^{n}$ for $n>3$
are absent in the effective action of a neutral scalar field at the
long wavelength limit, even the well-known dimensionless coupling
$\lambda\phi_{\star}^{6}$. In contrast, we find that only derivative
couplings are present in this situation.

Once again, similarly to the case of $\phi\phi$ contribution, we also see that a Chern--Simons-like term does not appear in the analysis of the $\phi\phi\phi$ contribution. This can be easily seen by considering the possible Chern--Simons-like expressions described by $\int d^{3}x~\varepsilon^{\mu\nu\beta}(\partial_{\mu}\phi)\star(\partial_{\nu}\phi)\star(\partial_{\beta}\phi)$ or
$\int d^{3}x~\varepsilon^{\mu\nu\beta}(\partial_{\mu}\partial_{\alpha}\phi)\star (\partial_{\nu}\partial^{\alpha}\phi)\star(\partial_{\beta}\phi)$ that are apparently nonzero but are in fact both of them vanish, due to the integration by part and discarding the surface terms.
%%%%%%%%%%%%%%%%%%%%%%%%%%%%%%%%%%%%%%%%%%%%%%%%%%%%%%%%%%%%%%%%%%%
\subsection{Charged scalar fields}
%%%%%%%%%%%%%%%%%%%%%%%%%%%%%%%%%%%%%%%%%%%%%%%%%%%%%%%%%%%%%%%%%%%
For completeness, in addition to the discussion of a neutral scalar
field, let us consider the case of charged scalar fields too. The
fermionic action in this case is defined as
\begin{equation}
S=\int d^{3}x\left[i\bar{\psi}\star
\gamma^{\mu}D_{\mu}^{\star}\psi-m\bar{\psi}\star\psi+g\bar{\psi}\star\Phi\star\psi+h.c.\right],\label{eq:2.1}
\end{equation}
in which the covariant derivative is defined as
$D_{\mu}^{\star}\psi=\partial\psi-ieA_{\mu}\star\psi$, and the h.c.
term ensures the reality of the action. Moreover, this action is
invariant under the local infinitesimal gauge transformations,
\begin{equation}
\delta\psi=ig\lambda\star\psi,\quad\delta
A_{\mu}=\partial_{\mu}\lambda-ie\left[A_{\mu},\lambda\right]_{\star}.
\end{equation}
The one-loop effective action coming from \eqref{eq:2.1} are
readily obtained,
\begin{equation}
i\Gamma\left[A\right]=-\sum_{n}\frac{1}{n}tr\bigg((\displaystyle{\not}\partial+im)^{-1}i\left(g\phi\star+e\displaystyle{\not}A\star\right)\bigg)^{n},\label{eq:2.2}
\end{equation}
where we have defined by simplicity the combination
$\phi=\Phi+\Phi^{\dagger}$. The $n=2$ contribution is exactly the
same as the one obtained in \eqref{eq:0.13}, just with the previous
replacement on the field $\phi$. We shall now proceed to analyse the
interacting terms between the scalar and gauge fields coming from
the $n=3$ and $n=4$ terms.

%%%%%%%%%%%%%%
\subsubsection{$\phi\phi A$ contribution}
\begin{figure}[t]
\vspace{-0.5cm}
\includegraphics[width=4.6cm,height=3.2cm]{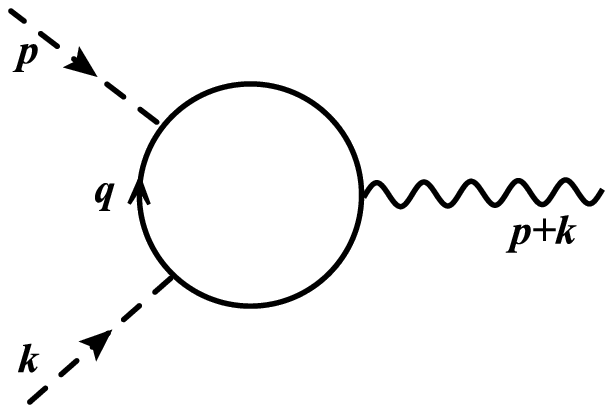}
 \centering  \caption{Relevant graph for the induced $\phi\phi A$-term.}
\label{oneloopdiagrams3}
\end{figure}

%%%%%%%%%%%%%%
Let us start with the first contribution coming from $n=3$ and depicted in Fig.~\ref{oneloopdiagrams3}.
Among all these interacting terms coming from this expansion we
shall concentrate in those giving a combination of $\phi\phi A$
fields. We thus find that three terms are present and have the
following structure
\begin{align}
i\Gamma\left[\phi\phi A\right] & \simeq\int d^{3}x_{1}d^{3}x_{2}d^{3}x_{3}
\biggl[[\phi\left(x_{1}\right)\phi\left(x_{2}\right)A_{\mu}\left(x_{3}\right)]\Gamma_{\left(a\right)}^{\mu}\left(x_{1},x_{2},x_{3}\right)\label{eq:2.4} \\
 & +[\phi\left(x_{1}\right)A_{\mu}\left(x_{2}\right)\phi\left(x_{3}\right)]\Gamma_{\left(b\right)}^{\mu}\left(x_{1},x_{2},x_{3}\right)
 +[A_{\mu}\left(x_{1}\right)\phi\left(x_{2}\right)\phi\left(x_{3}\right)]
 \Gamma_{\left(c\right)}^{\mu}\left(x_{1},x_{2},x_{3}\right)\biggr],\nonumber
\end{align}
where we define and compute the general tensor quantities $\Gamma_{\left(i\right)}^{\mu}\left(x_{1},x_{2},x_{3}\right) $, for $i=a,b,c$,  in the Appendix \ref{appB}.

%%%%%%%%%%%%%%%%%%%%%%%%%%%%%%%%%%%%%%%%%%%%%%%%%%%%
%%%%%%%%%%%%%%%%%%%%%%%%%%%%%%%%%%%%%%%%%%%%%%%%%%%%
%%%%%%%%%%%%%%%%%%%%%%%%%%%%%%%%%%%%%%%%%%%%%%%%%%%%
%%%%%%%%%%%%%%%%%%%%%%%%%%%%%%%%%%%%%%%%%%%%%%%%%%%%

Therefore, substituting the results \eqref{eq:2.6b} and \eqref{eq:2.6c}
back into the expression \eqref{eq:2.4}, and after some
straightforward integral manipulation we are finally able to write
\begin{align}
i\Gamma_{\left(a\right)}\left[\phi\phi A\right] & =\frac{g^{2}e}{36\pi}\frac{1}{\left|m\right|}\int d^{3}x\left[\partial^{\mu}\phi\star\phi\star A_{\mu}-\phi\star\partial^{\mu}\phi\star A_{\mu}\right],\label{eq:2.7a}\\
i\Gamma_{\left(b\right)}\left[\phi\phi A\right] & =-\frac{g^{2}e}{36\pi}\frac{1}{\left|m\right|}\int d^{3}x\left[\partial^{\mu}\phi\star A_{\mu}\star\phi-\phi\star A_{\mu}\star\partial^{\mu}\phi\right],\label{eq:2.7b}\\
i\Gamma_{\left(c\right)}\left[\phi\phi A\right] &
=\frac{g^{2}e}{36\pi}\frac{1}{\left|m\right|}\int
d^{3}x\left[A_{\mu}\star\partial^{\mu}\phi\star\phi-A_{\mu}\star\phi\star\partial^{\mu}\phi\right].
\label{eq:2.7c}
\end{align}
The complete contribution is found by summing the above three
contributions, Eqs.\eqref{eq:2.7a}--\eqref{eq:2.7c}. Thus, using the
cyclic property of the Moyal product, we find
\begin{align}
i\Gamma\left[\phi\phi A\right] & =i\Gamma_{\left(a\right)}\left[\phi\phi A\right]+i\Gamma_{\left(b\right)}\left[\phi\phi A\right]+i\Gamma_{\left(c\right)}\left[\phi\phi A\right],\nonumber \\
 & =\frac{g^{2}e}{12\pi}\frac{1}{\left|m\right|}\int d^{3}x\left[A_{\mu}\star\partial^{\mu}\phi\star\phi-A_{\mu}\star\phi\star\partial^{\mu}\phi\right].\label{eq:2.8}
\end{align}
At last, by using the definition
$\phi\rightarrow\Phi+\Phi^{\dagger}$ and keeping the relevant terms,
we can rewrite expression \eqref{eq:2.8} into the following
convenient form
\begin{align}
i\Gamma[\Phi\Phi^{\dagger} A] &
\simeq\frac{g^{2}e}{12\pi}\frac{1}{\left|m\right|}\int
d^{3}x\bigg(\left[\Phi^{\dagger},A_{\mu}\right]_{\star}\star\partial^{\mu}\Phi-\partial^{\mu}\Phi^{\dagger}\star
\left[A_{\mu},\Phi\right]_{\star}\bigg).\label{eq:2.9}
\end{align}

As we will see afterwards, this result is exactly the cubic
interaction from the NC Higgs model, since the coupling in this case
is given by $\left({\cal{D}}_{\mu}\Phi\right)^{\dagger}\star {\cal{D}}^{\mu}\Phi$, where the covariant derivative is now written in its adjoint form
${\cal{D}}_{\mu}\Phi=\partial_{\mu}\Phi
-ie\left[A_{\mu},\Phi\right]_{\star}$.
It is worth of mention that the generated couplings are in the adjoint representation and not in the fundamental one.\footnote{This fact might be closely related to our choice of interaction term in \eqref{eq:0.1}, perhaps because different couplings such as $ \bar\psi\star\psi\star\phi$ or $\bar\psi\star[\phi,\psi]_\star $ could give in principle different contributions to the effective action. However, this fact should be further elaborated and then analysed.}
%%%%%%%%%%%%%%%%%%%%%%%%%%%%%%%%%%%%%%%%%%%%%%%%%%%%
%%%%%%%%%%%%%%%%%%%%%%%%%%%%%%%%%%%%%%%%%%%%%%%%%%%%
\begin{figure}[t]
\vspace{-0.5cm}
\includegraphics[width=4.5cm,height=3.3cm]{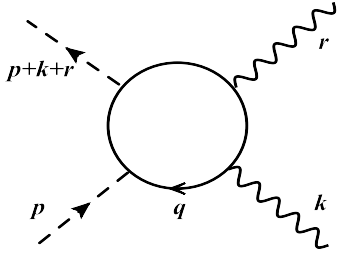}
 \centering \caption{Relevant graph for the induced $\phi\phi AA$-term.}
\label{oneloopdiagrams4}
\end{figure}
%%%%%%%%%%%%%%%%%%%%%%%%%%%%%%%%%%%%%%%%%%%%%%%%%%%%
%%%%%%%%%%%%%
\subsubsection{$\phi\phi AA$ contribution}
%%%%%%%%%%%%%
By means of complementarity, we now approach the one-loop $n=4$
contribution to the charged scalar fields effective action, which graph is given in Fig.~\ref{oneloopdiagrams4}, where we
shall find the last piece of the interacting sector.

The contribution proportional to the following structure of
$\phi\phi AA$ fields is given by six terms
\begin{align}
i\Gamma\left[\phi\phi AA\right] & \simeq\int d^{3}x_{1}\cdots
d^{3}x_{4}\biggl[[\phi\left(x_{1}\right)\phi\left(x_{2}\right)A_{\mu}
\left(x_{3}\right)A_{\nu}\left(x_{4}\right)]\Gamma_{\left(a\right)}^{\mu\nu}
 +[\phi\left(x_{1}\right)A_{\mu}\left(x_{2}\right)\phi\left(x_{3}\right)A_{\nu}\left(x_{4}\right)]\Gamma_{\left(b\right)}^{\mu\nu} \nonumber \\
 & +[\phi\left(x_{1}\right)A_{\mu}\left(x_{2}\right)A_{\nu}\left(x_{3}\right)\phi\left(x_{4}\right)]\Gamma_{\left(c\right)}^{\mu\nu}
 +[A_{\mu}\left(x_{1}\right)\phi\left(x_{2}\right)\phi\left(x_{3}\right)A_{\nu}\left(x_{4}\right)]\Gamma_{\left(d\right)}^{\mu\nu} \nonumber \\
 & +[A_{\mu}\left(x_{1}\right)\phi\left(x_{2}\right)A_{\nu}\left(x_{3}\right)\phi\left(x_{4}\right)]\Gamma_{\left(e\right)}^{\mu\nu}
 +[A_{\mu}\left(x_{1}\right)A_{\nu}\left(x_{2}\right)\phi\left(x_{3}\right)\phi\left(x_{4}\right)]\Gamma_{\left(f\right)}^{\mu\nu} \biggr],\label{eq:2.10}
\end{align}
 where we define and compute the tensor quantities
to each one of these contributions $\Gamma_{\left(i\right)}^{\mu\nu}$ $i=a,b,...,f$, in the Appendix~\ref{appB}.

Hence, replacing the results \eqref{eq:2.14b} in conjunction with \eqref{eq:2.12a} and
\eqref{eq:2.12b} back into the expression \eqref{eq:2.10}, we
find out the following
\begin{align}
i\Gamma\left[\phi\phi AA\right]= &
\frac{i}{12\pi}\frac{g^{2}e^2}{\left|m\right|}\int
d^{3}x\left[\phi\star A_{\mu}\star A^{\mu}\star\phi-\phi\star
A_{\mu}\star\phi\star A^{\mu}\right].\label{eq:2.15}
\end{align}
Now, making use again of $\phi\rightarrow\Phi+\Phi^{\dagger}$, we
finally find the expression for the effective action
\begin{align}
i\Gamma\left[\Phi\Phi^{\dagger} AA\right] &
=\frac{i}{12\pi}\frac{g^{2}e^2}{\left|m\right|}\int
d^{3}x\bigg(\left[\Phi^{\dagger},A_{\mu}\right]_{\star}\star\left[A^{\mu},\Phi\right]_{\star}\bigg).\label{eq:2.16}
\end{align}

As we have anticipated, the expression \eqref{eq:2.16} is precisely the
last piece for the (minimal) interaction content of the NC Higgs model
$\left({\cal{D}}_{\mu}\Phi\right)^{\dagger}\star {\cal{D}}^{\mu}\Phi$, consisting in
the quartic interaction among the scalar and gauge fields.

%%%%%%%%%%%%%%%%%%%%%%%%%%%%%%%%%%%%%%%%%%%%%%%%%%%%%
%%%%%%%%%%%%%%%%%%%%%%%%%%%%%%%%%%%%%%%%%%%%%%%%%%%%%
\section{Propagating modes scalar field}
\label{sec4}
%%%%%%%%%%%%%%%%%%%%%%%%%%%%%%%%%%%%%%%%%%%%%%%%%%%%%
%%%%%%%%%%%%%%%%%%%%%%%%%%%%%%%%%%%%%%%%%%%%%%%%%%%%%

From the obtained results, Eqs.\eqref{eq:0.13} and \eqref{eq: 1.12},
we can analyse the dynamics of the scalar fields by proposing the following effective Lagrangian for the
noncommutative neutral scalar field
\begin{equation}
\mathcal{L}=\frac{1}{2}\left(\partial_{\mu}\phi\partial^{\mu}\phi-m^{2}\phi^{2}\right)
+\frac{\lambda}{3!}\left(\phi\star\phi\star\phi+\frac{1}{3m^{2}}\partial^{\mu}\phi\star\partial_{\mu}\phi\star\phi\right).\label{eq:8.1}
\end{equation}
Notice that the usual $\phi^3_\star$ theory is recovered in the limit when the HD contribution decouples.
The Feynman rules for this theory are readily obtained from the Lagrangian \eqref{eq:8.1}.

%\begin{itemize}
 % \item The scalar
%field propagator
%\begin{equation}
%S\left(p\right)=\frac{i}{p^{2}-m^{2}}.
%\end{equation}
%%
%  \item  The cubic scalar field vertex
%\begin{align}
%\Gamma\left(p_{1},p_{2}\right) &
%=\sqrt{2}~\lambda\biggl[1+\frac{1}{9m^{2}}\left(\left(p_{1}\right)^{2}
%+p_{1}.p_{2}+(p_{2})^{2}\right)\biggr]\cos\left[\frac{p_{1}\times p_{2}}{2}\right].
%\end{align}
%\end{itemize}

\begin{figure}[t]
\vspace{-0.5cm}
\includegraphics[width=5cm,height=2.5cm]{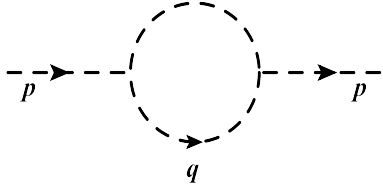}
 \centering \protect\protect\protect\caption{One-loop self-energy graph for the neutral scalar field.}
\label{oneloopdiagrams5}
\end{figure}

Moreover, we can establish the renormalization of the complete
propagator, by writing the self-energy function as
$\Sigma\left(p^{2}\right)=p^{2}\Sigma_{1}\left(p^{2}\right)+m^{2}\Sigma_{2}\left(p^{2}\right)$,
that can be carried out as
\[
\mathcal{S}\left(p\right)=\frac{1}{p^{2}-m^{2}-\Sigma\left(p^{2}\right)}=\frac{1}{p^{2}\left(1-\Sigma_{1}\left(p^{2}\right)\right)-m^{2}\left(1+\Sigma_{2}\left(p^{2}\right)\right)}=\frac{\mathcal{Z}}{p^{2}-m_{ren}^{2}}
\]
where we have defined the renormalization constants as
$\mathcal{Z}^{-1}=1-\Sigma_{1}\left(p^{2}\right)$ and
$\mathcal{Z}_{m}^{-1}=1+\Sigma_{2}\left(p^{2}\right)$, so that the
renormalized mass is defined as the following
\begin{equation}
m_{ren}=m\sqrt{\frac{\mathcal{Z}}{\mathcal{Z}_{m}}}.\label{eq:8.8}
\end{equation}
Hence, the one-loop self-energy for the neutral scalar fields reads  (see Fig.~\ref{oneloopdiagrams5})
\begin{align}
\Sigma\left(p\right) &  = -i\lambda^{2}\int\frac{d^{3}q}{\left(2\pi\right)^{3}}\frac{1}{q^{2}-m^{2}}
 \frac{1}{\left(q-p\right)^{2}-m^{2}}\left[1+\frac{1}
 {9m^{2}}\left(\left(p.q\right)+\left(q-p\right)^{2}\right)\right]^{2}
 \cos^{2}\left[\frac{p\times q}{2}\right],\label{eq:8.2}
\end{align}
We can work out the numerator of the above expression in order to simplify the
dependence on the integrated momentum.
Furthermore, we now compute separately the planar from the
non-planar contribution by means of the identity
$2\cos^{2}[\frac{p\times Q}{2}]=1+\cos\left(p\times Q\right)$. First,
the planar contribution results into the following
\begin{align}
\Sigma_{p}\left(p\right) & =\frac{1}{1296\pi}\frac{\lambda^{2}}{m}\biggl\{\int_{0}^{1}dx\left[100+20x\beta+x^{2}\beta^{2}\right]
\frac{1}{\left(1-x\left(1-x\right)\beta\right)^{\frac{1}{2}}}\nonumber \\
 & +\beta\int_{0}^{1}dx\sqrt{1-x\left(1-x\right)\beta}+2\left[20+\beta\right]\biggr\},\label{eq:8.4}
\end{align}
where we have defined the notation $\beta=p^{2}/m^{2}$. Next, the
non-planar contribution from \eqref{eq:8.2} can be computed with
help of the results Eqs.\eqref{eq:C.1} and \eqref{eq:C.2}, and
yields
\begin{align}
\Sigma_{n-p}\left(p\right) & =\frac{1}{1296\pi}\frac{\lambda^{2}}{m}\biggl\{\int_{0}^{1}dx\left[100+20x\beta+x^{2}\beta^{2}\right]
\frac{e^{-m\left|\tilde{p}\right|\sqrt{1-x\left(1-x\right)\beta}}}{\sqrt{1-x\left(1-x\right)\beta}}
\nonumber \\
 & -\frac{\beta}{m\left|\tilde{p}\right|}\int_{0}^{1}dx~
 e^{-m\left|\tilde{p}\right|\sqrt{1-x\left(1-x\right)\beta}}-2\left[20+\beta\right]
 \frac{1}{m\left|\tilde{p}\right|}e^{-m\left|\tilde{p}\right|}\biggr\}.\label{eq:8.6}
\end{align}
Finally, the complete contribution is given by the sum of
\eqref{eq:8.4} and \eqref{eq:8.6},
$\Sigma\left(p\right)=\Sigma_{p}\left(p\right)+\Sigma_{n-p}\left(p\right)$.

Now with the one-loop self-energy we can analyse the renormalized
mass expression structure. Thus, expanding \eqref{eq:8.8} at leading order, we
find that
\begin{equation}
m_{ren}=m\sqrt{\frac{\mathcal{Z}}{\mathcal{Z}_{m}}}\simeq
m+\Sigma^{\left(1\right)}+\mathcal{O}\left(\alpha^{2}\right),
\end{equation}
where we have defined
$\Sigma^{\left(1\right)}=\frac{m}{2}\left(\Sigma_{1}\left(m^{2}\right)+\Sigma_{2}\left(m^{2}\right)\right)$,
with the previous coefficients given by
\begin{align}
\Sigma_{1}\left(m^{2}\right) & =\frac{1}{1296\pi}\frac{\lambda^{2}}{m^{3}}\biggl\{\int_{0}^{1}dx\frac{\left[20x+x^{2}\right]}
{\sqrt{1-x\left(1-x\right)}}\left[1+e^{-m\left|\tilde{p}\right|\sqrt{1-x\left(1-x\right)}}\right]\nonumber \\
 & +\int_{0}^{1}dx\left[\sqrt{1-x\left(1-x\right)}-\frac{1}{m\left|\tilde{p}\right|}
 e^{-m\left|\tilde{p}\right|\sqrt{1-x\left(1-x\right)}}\right]
 +2\left(1-\frac{1}{m\left|\tilde{p}\right|}e^{-m\left|\tilde{p}\right|}\right)\biggr\}.
\end{align}
and
\begin{align}
\Sigma_{2}\left(m^{2}\right) &
=\frac{1}{1296\pi}\frac{\lambda^{2}}{m^{3}}\biggl\{\int_{0}^{1}dx\frac{100}{\sqrt{1-x\left(1-x\right)}}
\left[1+e^{-m\left|\tilde{p}\right|\sqrt{1-x\left(1-x\right)}}\right]
+40\left(1-\frac{1}{m\left|\tilde{p}\right|}e^{-m\left|\tilde{p}\right|}\right)\biggr\}.
\end{align}
Hence, we see that the dispersion relation for the scalar field at
this order reads
\begin{equation}
\omega^{2}=\vec{p}^{2}+m_{ren}^{2}\simeq\vec{p}^{2}+m^{2}+2m\Sigma^{\left(1\right)}+\mathcal{O}\left(\alpha^{2}\right).\label{eq:8.10}
\end{equation}
By means of illustration, we shall consider the infinitesimal
noncommutative modification in the dispersion relation. Thus, if we
additionally take the on-shell limit, i.e.
$\left|\tilde{p}\right|=\left|\theta\right|\sqrt{p^{2}}\rightarrow
m\left|\theta\right|$, we can write the dispersion relation in the
following simple form
\begin{equation}
\omega^{2}\simeq\vec{p}^{2}+m^{2}+\frac{1}{1296\pi}\frac{\lambda^{2}}{m}\left(86+\frac{441}{2}\log3\right)-\frac{43}{1296\pi}\frac{\lambda^{2}}{m^{3}\theta}+\mathcal{O}\left(\lambda^{2}\theta\right).\label{eq:8.11}
\end{equation}
Immediately we observe two features from the expression
\eqref{eq:8.11}. First, we find a correction for the mass as
$m_{eff}^{2}=m^{2}\left[1+\frac{\lambda^{2}}{\pi
m^{3}}\left(\frac{43}{648}+\frac{49}{288}\log3\right)\right]$,
meaning that the particle gets heavier. Second, the last term shows
the presence of a UV/IR instability (with a $1/\theta$
behavior) caused by noncommutative perturbative effects.

%%%%%%%%%%%%%%%%%%%%%%%%%%%%%%%%%%%%%%%%%%%%%%%%%%%%%%%%%%%
%%%%%%%%%%%%%%%%%%%%%%%%%%%%%%%%%%%%%%%%%%%%%%%%%%%%%%%%%%
\section{Propagating modes charged scalar fields}
\label{sec5}
%%%%%%%%%%%%%%%%%%%%%%%%%%%%%%%%%%%%%%%%%%%%%%%%%%%%%%%%%%%
%%%%%%%%%%%%%%%%%%%%%%%%%%%%%%%%%%%%%%%%%%%%%%%%%%%%%%%%%%

In addition, we now consider the obtained results \eqref{eq:2.9} and
\eqref{eq:2.16} (the gauge field effective action was obtained in \cite{bufalo}), so that we can propose the following effective
Lagrangian for the charged scalar fields coupled
with a higher derivative Chern-Simons field
\begin{align}
\mathcal{L} & =\left({\cal{D}}_{\mu}\Phi\right)^{\dagger}\star {\cal{D}}^{\mu}\Phi-m^{2}\Phi^{\dagger}\star\Phi+\frac{m}{2}\varepsilon^{\mu\nu\sigma}A_{\mu}\left(1+\frac{\Box}{m^{2}}\right)\partial_{\nu}A_{\sigma}\nonumber \\
 & -\frac{1}{2\xi}\left(\partial_{\mu}A^{\mu}\right)^{2}+\frac{me}{3}\varepsilon^{\mu\nu\sigma}A_{\mu}\star A_{\nu}\star A_{\sigma}+\partial^{\mu}\overline{c}\star D_{\mu}c,\label{eq:8.12}
\end{align}
where the covariant derivatives are defined as such
$ {\cal{D}}_{\mu}=\partial_{\mu}-ie\left[A_{\mu},~\right]_{\star}$.
Based on the Lagrangian \eqref{eq:8.12} we will study the dynamics of the scalar and gauge fields.

Notice that the model described by the Lagrangian \eqref{eq:8.12}
has a non-gauge invariant contribution, given by the higher
derivative (HD) term. The
generation of higher derivative terms and derivative couplings within
the NC Chern-Simons theory were considered in Ref.~\cite{bufalo}
Moreover, our interest in exploring the features of this HD term into
the propagator is motivated by the possibility of finding
non-trivial effects into the whole NC Chern-Simons-Higgs theory
\eqref{eq:8.12}. For instance, as we will shortly show (see
Sect.\ref{sec5.2}) this solely HD contribution is responsible for
obtaining nontrivial outcomes for the pure NC CS theory (which is a
free theory without the HD term), so its effect on the complete
theory is expected to be rather interesting. Moreover, it is easy to
see that the commutative limit of this NC HD Chern-Simons theory is
also a free theory.

%The related Feynman rules for this NC HD-Chern-Simons-Higgs theory are as follows:

%\begin{itemize}
 % \item The scalar field propagator
%\begin{equation}
%S\left(p\right)=\frac{i}{p^{2}-m^{2}}.
%\end{equation}
%%
 % \item The gauge field propagator
%\begin{equation}
%iD_{\mu\nu}\left(k\right)=m\frac{i\varepsilon_{\mu\nu\sigma}k^{\lambda}}{k^{2}
%\left(k^{2}-m^{2}\right)}+\xi\frac{k_{\mu}k_{\nu}}{k^{4}}.\label{eq:8.12b}
%\end{equation}
%%
%  \item The ghost fields propagator
%\begin{equation}
%D\left(p\right)=\frac{i}{p^{2}}.
%\end{equation}
%%
%\item The cubic gauge vertex
%\begin{align}
%\Gamma^{\mu\nu\sigma}\left(p_{1},p_{2}\right) &
%=-i\sqrt{2}~me\varepsilon^{\mu\nu\sigma}
%\sin[\frac{p_{1}\times p_{2}}{2}].\label{eq:6.1}
%\end{align}
%%
%\item The cubic interaction among the ghost and gauge fields
%\begin{align}
%\Upsilon^{\mu}\left(p_{1},p_{2}\right) &
%=-i\sqrt{2}~ep_{1}^{\mu}\sin[\frac{p_{1}\times p_{2}}{2}].\label{eq:6.2}
%\end{align}
%%
%\item The cubic interaction of the charged scalar and gauge fields
%\begin{align}
%\Gamma^{\mu}\left(p_{1},p_{2}\right) &
%=-i\sqrt{2}~e\left(p_{1}+2p_{2}\right)^{\mu}\sin[\frac{p_{1}\times p_{2}}{2}].\label{eq:6.3}
%\end{align}
%%
%\item The quartic interaction among the charged scalar and gauge fields,
%\begin{eqnarray}
%\Gamma^{\mu\nu}\left(p_{1},p_{2},p_{3},p_{4}\right)  &=&-i\left(\sqrt{2}\right)^{2}e^{2}\eta^{\mu\nu}\bigg(\sin[\frac{p_{1}\times p_{3}}{2}]\sin[\frac{p_{2}\times p_{4}}{2}]\nonumber\\
%&+&\sin[\frac{p_{2}\times p_{3}}{2}]\sin[\frac{p_{1}\times p_{4}}{2}]\bigg).\label{eq:6.4}
%\end{eqnarray}
%\end{itemize}

We shall now approach two physical situations in the NC HD-Chern-Simons-Higgs
theory: we consider first the dispersion relation of the scalar
fields, where the two-point function renormalization takes place as before; second, we
analyse the dispersion relation for the gauge field.

\subsection{Dispersion relation charged scalar fields}
\label{sec5.1}

%%%%%%%%%%%%%%%%%%%%%%%%%%%%%%%%%%%%%%%%%%%%%%%%%%%%%%%%%%%%%%%%%%
%%%%%%%%%%%%%%%%%%%%%%%%%%%%%%%%%%%%%%%%%%%%%%%%%%%%%%%%%%%%%%%%%%
\begin{figure}[t]
\vspace{-2.5cm}
\includegraphics[width=8cm,height=5cm]{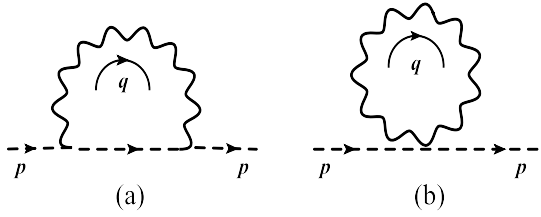}
 \centering  \caption{ One-loop self-energy graphs for the charged scalar field: (a) cubic interaction contribution, (b) quartic interaction contribution.}
\label{oneloopdiagrams6}
\end{figure}
%%%%%%%%%%%%%%%%%%%%%%%%%%%%%%%%%%%%%%%%%%%%%%%%%%%%%%%%%%%%%%%%%%
%%%%%%%%%%%%%%%%%%%%%%%%%%%%%%%%%%%%%%%%%%%%%%%%%%%%%%%%%%%%%%%%%%

For the one-loop self-energy contribution, we have the following two diagrams (Fig.~\ref{oneloopdiagrams6}), the whole contribution reads
\begin{align}
\Sigma\left(p\right) & =\Sigma^{\textrm{(a)}}\left(p\right) +\Sigma^{\textrm{(b)}}\left(p\right), \nonumber \\
 & =\left(\sqrt{2}\right)^{2}ie^{2}\int\frac{d^{3}q}{\left(2\pi\right)^{3}} iD_{\mu\nu}\left(q\right) \biggl[\frac{\left(2p-q\right)^{\mu}\left(2p-q\right)^{\nu}}
 {\left(q-p\right)^{2}-m^{2}}-2\eta^{\mu\nu}\biggr]
 \sin^{2}\left[\frac{p\times q}{2}\right].\label{eq:8.13}
\end{align}
An important remark follows from \eqref{eq:8.13}, we observe that only
the symmetric part of the gauge field propagator
\begin{equation}
iD_{\mu\nu}\left(k\right)=m\frac{i\varepsilon_{\mu\nu\sigma}k^{\lambda}}{k^{2}
\left(k^{2}-m^{2}\right)}+\xi\frac{k_{\mu}k_{\nu}}{k^{4}}.\label{eq:8.12b}
\end{equation}
contributes at this perturbative order. Hence, only
a non-physical gauge-dependent contribution is present in the case
of a Chern-Simons gauge field. By completeness, we will discuss
below the result for the NC Maxwell-Chern-Simons theory.

Nonetheless, making use of the propagator \eqref{eq:8.12b} and
rewriting \eqref{eq:8.13} with help of Feynman parametrization, we
obtain
\begin{align}
\Sigma\left(p\right) & =2i\xi e^{2}\mu^{3-\omega}\int\frac{d^{\omega}Q}{\left(2\pi\right)^{\omega}}\biggl[8\int_{0}^{1}dx\left(1-x\right)\left(\left(p.Q\right)^{2}+x^{2}p^{4}\right)\frac{1}{\left(Q^{2}+x\left(1-x\right)p^{2}-xm^{2}\right)^{3}}\nonumber \\
 & -4\int_{0}^{1}dx\frac{xp^{2}}{\left(Q^{2}+x\left(1-x\right)p^{2}-xm^{2}\right)^{2}}
 +\frac{1}{Q^{2}-m^{2}}-\frac{2}{Q^{2}}\biggr]
 \sin^{2}\left[\frac{p\times Q}{2}\right].\label{eq:8.14}
\end{align}

Now, it is convenient to compute separately the planar from the
non-planar contributions from \eqref{eq:8.14}, this is achieved by means of
$2\sin^{2}[\frac{p\times Q}{2}]=1-\cos \left(p\times Q\right) $. We have,
for the planar contribution
\begin{align}
\Sigma_{p}\left(p\right) & =-\frac{\xi e^{2}m}{8\pi}\biggl[2\beta\int_{0}^{1}dx\left(1-x\right)\left[\frac{1}{\left(x -x\left(1-x\right)\beta\right)^{\frac{1}{2}}}-x^{2}\beta\frac{1}{\left(x -x\left(1-x\right)\beta\right)^{\frac{3}{2}}}\right]\nonumber \\
 & -4\beta\int_{0}^{1}dx~x\frac{1}{\left(x -x\left(1-x\right)\beta\right)^{\frac{1}{2}}}+2\biggr],\label{eq:8.15}
\end{align}
whereas, with help of \eqref{eq:C.1} and \eqref{eq:C.2}, the
non-planar contribution reads
\begin{align}
\Sigma_{n-p}\left(p\right) & =\frac{\xi e^{2}m}{4\pi}\biggl[-2\beta\int_{0}^{1}dxx\frac{e^{-m\left|\tilde{p}\right|\sqrt{x -x\left(1-x\right)\beta}}}{\sqrt{x -x\left(1-x\right)\beta}} -\frac{1}{m\left|\tilde{p}\right|}e^{-m\left|\tilde{p}\right|}+\frac{2}{m\left|\tilde{p}\right|} \nnr
&+ \beta\int_{0}^{1}dx\left(1-x\right)\left(1-x^{2}\beta\frac{\left(1+m\left|\tilde{p}\right|\sqrt{x -x\left(1-x\right)\beta}\right)}{\left(x -x\left(1-x\right)\beta\right)}\right)\frac{e^{-m\left|\tilde{p}\right|\sqrt{x -x\left(1-x\right)\beta}}}{\sqrt{x -x\left(1-x\right)\beta}}\biggr].\label{eq:8.16}
\end{align}
The renormalization here follows closely the steps as of the neutral
scalar field, so that the renormalized mass is also given by
\eqref{eq:8.8}. Hence, we can proceed and decompose the one-loop
self-energy in terms of its components,
$\Sigma\left(p^{2}\right)=p^{2}\Sigma_{1}\left(p^{2}\right)+m^{2}\Sigma_{2}\left(p^{2}\right)$.
Now, since the renormalization is performed on-shell, we have that these
components are simply written as
\begin{align}
\Sigma_{1}(m^{2}) & =-\frac{\xi e^{2}}{4\pi}\frac{1}{m}\biggl[\frac{1}{m\left|\tilde{p}\right|}-1-\frac{e^{-m\left|\tilde{p}\right|}}{m\left|\tilde{p}\right|}\biggr],\label{eq:8.17a}\\
\Sigma_{2}(m^{2}) & =-\frac{\xi
e^{2}}{4\pi}\frac{1}{m}\biggl[1+\frac{1}{m\left|\tilde{p}\right|}e^{-m\left|\tilde{p}\right|}-\frac{2}{m\left|\tilde{p}\right|}\biggr].\label{eq:8.17b}
\end{align}
As we have already discussed, the renormalized mass is given by
\eqref{eq:8.8}
$m_{ren}=m\sqrt{\frac{1+\Sigma_{2}(m^{2})}{1-\Sigma_{1}(m^{2})}}$,
so the dispersion relation for the charged scalar fields are written
as
\begin{equation}
\omega^{2}=\vec{p}^{2}+m_{ren}^{2}\simeq\vec{p}^{2}+m^{2}+m^{2}\left(\Sigma_{1}(m^{2})+\Sigma_{2}(m^{2})\right)+\mathcal{O}\left(\alpha^{2}\right).
\end{equation}
Finally, making use of \eqref{eq:8.17a} and \eqref{eq:8.17b}, and
recalling that at the on-shell limit we have
$\left|\tilde{p}\right|\rightarrow m\left|\theta\right|$, we thus
find
\begin{equation}
\omega^{2}\simeq\vec{p}^{2}+m^{2}+\frac{\xi
e^{2}}{4\pi}\frac{1}{m\theta}+\mathcal{O}\left(e^{2}\theta\right).\label{eq:8.18}
\end{equation}
Some conclusions can be depicted from the expression \eqref{eq:8.18}. We
first realize that in this framework no radiative correction for the
mass is found. However, we still have a UV/IR instability caused by
NC effects, in particular we see that this UV/IR instability is
proportional to the gauge parameter $\xi$, and for the Landau gauge,
this instability vanishes. We therefore conclude that this sector of the
theory is empty of physical content.

In order to illustrate the physical content of the scalar sector,
let us consider the usual Maxwell-Chern-Simons propagator (at Landau
gauge, $\xi=0$)\footnote{This is justifiable as an example, since we
have a symmetric operator multiplying the gauge field propagator in
\eqref{eq:8.13}, so the skew-symmetric nature of the pure
Chern-Simons propagator gives a vanishing result.}
\begin{equation}
iD_{\mu\nu}\left(k\right)=\frac{1}{k^{2}\left(k^{2}-m^{2}\right)}\left(k^{2}\eta_{\mu\nu}-k_{\mu}k_{\nu}+im\varepsilon_{\mu\nu\lambda}k^{\lambda}\right),\label{eq:8.19}
\end{equation}
this gives the following expression for the scalar field dispersion
relation
\begin{equation}
\omega^{2}\simeq\vec{p}^{2}+m^{2}-\frac{e^{2}m}{\pi}\left(1+\frac{e^{-m^{2}\theta}}{m^{2}\theta}\right)
+\mathcal{O}\left(e^{2}\theta\right),\label{eq:8.20}
\end{equation}
where we realize that in this framework of a Maxwell-Chern-Simons gauge
field, a radiative correction for the mass of the scalar field is
found, so that the effective mass reads
$m_{eff}^{2}=m^{2}\left[1-\frac{1}{\pi}\frac{e^{2}}{m}\right]$. In
the same way, we also have a UV/IR instability caused by NC effects.
All these effects present in \eqref{eq:8.20} are caused by the
symmetric component of the propagator \eqref{eq:8.19}.

\subsection{Dispersion relation gauge field}
\label{sec5.2}

In general, we can consider that the self-energy $1PI$-function in a three dimensional noncommutative spacetime has the following
tensor form \cite{ghasemkhani}
\begin{align}
\Pi^{\mu\nu}= &
\bigg(\eta^{\mu\nu}-\frac{p^{\mu}p^{\nu}}{p^{2}}\bigg)\Pi_{\textrm{e}}
+\frac{\tilde{p}^{\mu}\tilde{p}^{\nu}}{\tilde{p}^{2}}~\widetilde{\Pi}_{\textrm{e}}
+i\Pi_{\textrm{0}}^{\textrm{\tiny{A}}}\epsilon^{\mu\nu\lambda}p_{\lambda}+\Pi_{\textrm{0}}^{\textrm{\tiny{S}}}
\bigg(\tilde{p}^{\mu}u^{\nu}+\tilde{p}^{\nu}u^{\mu}\bigg),\label{eq:7.1}
\end{align}
where we regard the basis composed by the vectors $p^{\mu}$, $\tilde{p}^{\mu}$
and $u_{\mu}=\epsilon_{\mu\alpha\beta}p^{\alpha}\tilde{p}^{\beta}$;
moreover, the form factors $\Pi_{\textrm{e}}$,
$\widetilde{\Pi}_{\textrm{e}}$,
$\Pi_{\textrm{0}}^{\textrm{\tiny{A}}}$ and
$\Pi_{\textrm{0}}^{\textrm{\tiny{S}}}$ are
determined by means of the following relations
\begin{align}
\Pi_{\textrm{e}}= & \eta_{\mu\nu}\Pi^{\mu\nu}-\frac{\tilde{p}_{\mu}\tilde{p}_{\nu}}{\tilde{p}^{2}}\Pi^{\mu\nu},\label{eq:7.2a}\\
\widetilde{\Pi}_{\textrm{e}}= & -\eta_{\mu\nu}\Pi^{\mu\nu}+2\frac{\tilde{p}_{\mu}\tilde{p}_{\nu}}{\tilde{p}^{2}}\Pi^{\mu\nu},\label{eq:7.2b}\\
\Pi_{\textrm{0}}^{\textrm{\tiny{A}}}= & \frac{i}{2p^{2}}\epsilon_{\mu\nu\alpha}p^{\alpha}\Pi^{\mu\nu},\label{eq:7.2c}\\
\Pi_{\textrm{0}}^{\textrm{\tiny{S}}}= &
-\frac{1}{2\tilde{p}^{4}p^{2}}\left(u_{\mu}\tilde{p}_{\nu}+u_{\nu}\tilde{p}_{\mu}\right)\Pi^{\mu\nu}.\label{eq:7.2d}
\end{align}
Moreover, the general expression of the complete propagator for the
 CS gauge field (augmented with the HD term) can be put into the form
\cite{ghasemkhani}
\begin{align}
i\mathcal{D}_{\mu\nu} & =-\frac{\Pi_{\textrm{e}}+\widetilde{\Pi}_{\textrm{e}}}{D}\eta_{\mu\nu}+\left(\frac{\Pi_{\textrm{e}}
+\widetilde{\Pi}_{\textrm{e}}}{D}+\frac{\xi}{p^{2}}\right)\frac{p_{\mu}p_{\nu}}{p^{2}}
+\frac{\widetilde{\Pi}_{\textrm{e}}}{D}\frac{\tilde{p}_{\mu}\tilde{p}_{\nu}}{\tilde{p}^{2}}\nonumber \\
 & +\frac{\Pi_{\textrm{0}}^{\textrm{\tiny{S}}}}{D}\left(\tilde{p}_{\mu}u_{\nu}+u_{\mu}\tilde{p}_{\nu}\right)
 +\frac{m\left(1-\frac{p^{2}}{m^{2}}\right)+\Pi_{\textrm{0}}^{\textrm{\tiny{A}}}}{D}
 i\varepsilon_{\mu\nu\lambda}p^{\lambda}.\label{eq:7.3}
\end{align}
where by simplicity we have defined the quantity at the denominator
\begin{equation}
D=\Pi_{\textrm{e}}\left(\Pi_{\textrm{e}}+\widetilde{\Pi}_{\textrm{e}}\right)
+p^{2}\left[\left(\tilde{p}^{2} \Pi_{\textrm{0}}^{\textrm{\tiny{S}}}\right)^{2}-\left(m\left(1-\frac{p^{2}}{m^{2}}\right)
+\Pi_{\textrm{0}}^{\textrm{\tiny{A}}}\right)^{2}\right].
\end{equation}
Before discussing the structure of the complete propagator we shall
now compute the one-loop contribution to the polarization tensor.
%%%%%%%%%%%%%%%%%%%%%%%%%%%%55
\begin{figure}[t]
\vspace{-0.5cm}
\includegraphics{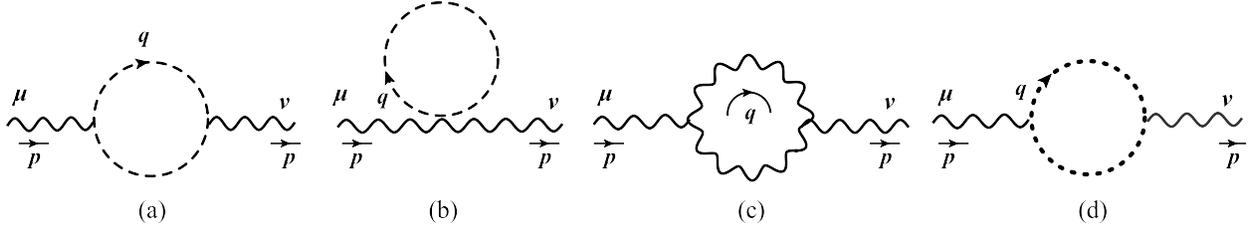}
 \centering \caption{ One-loop self-energy graphs for the gauge field: (a) scalar loop, (b) scalar tadpole loop, (c) gauge loop, (d) ghost loop.}
\label{oneloopdiagrams7}
\end{figure}
%%%%%%%%%%%%%%%%%%%%%%%%%%%

The one-loop correction to the gauge field self-energy is given by
the four contributions depicted in Fig.~\ref{oneloopdiagrams7}:
the first and second contributions are from the scalar loops, graph (a) corresponds
to the scalar loop , while graph (b) to the scalar tadpole loop ,
graph (c) corresponds to the gauge loop , and graph (d) corresponds to the ghost loop, respectively.
 Their explicit expressions are written as
\begin{align}
\Pi_{\mu\nu}^{\textrm{(a+b)}}\left(p\right) &
=-2ie^{2}\int\frac{d^{3}q}{\left(2\pi\right)^{3}}\frac{1}{q^{2}-m^{2}}
\left[\frac{\left(2q-p\right)^{\mu}\left(2q-p\right)^{\nu}}
{\left(q-p\right)^{2}-m^{2}}-2\eta^{\mu\nu}\right]
\sin^{2}\left[\frac{p\times q}{2}\right],\label{eq:8.21a}\\
\Pi_{\mu\nu}^{\textrm{(c)}}\left(p\right) & =-im^{4}e^{2}\int\frac{d^{3}q}{\left(2\pi\right)^{3}}\frac{q^{\nu}\left(q-p\right)^{\mu}+q^{\mu}
\left(q-p\right)^{\nu}}{\left(q-p\right)^{2}\left(\left(q-p\right)^{2}-m^{2}\right)q^{2}
\left(q^{2}-m^{2}\right)}
\sin^{2}\left[\frac{p\times q}{2}\right],\label{eq:8.21b}\\
\Pi_{\mu\nu}^{\textrm{(d)}}\left(p\right) & =2ie^{2}\int\frac{d^{3}q}{\left(2\pi\right)^{3}}\frac{1}{q^{2}}\frac{\left(q-p\right)^{\mu}q^{\nu}}
{\left(q-p\right)^{2}}\sin^{2}\left[\frac{p\times q}{2}\right]
.\label{eq:8.21c}
\end{align}
In the first line, by simplicity of notation, we have summed the two
contributions coming from the scalar loops.

An interesting remark is now in place. In particular, in the limit
$m\rightarrow\infty$, that corresponds to the situation where the HD contribution is removed,
we find that the scalar contributions \eqref{eq:8.21a} are equal to
zero, while the contributions \eqref{eq:8.21b} and \eqref{eq:8.21c}
sum to
\begin{align}
\Pi_{\mu\nu}^{\textrm{(c+d)}}\left(p\right)&
=ie^{2}\int_{0}^{1}dy\int\frac{d^{3}Q}{\left(2\pi\right)^{3}}\frac{\left[Q^{\nu}p^{\mu}-p^{\nu}Q^{\mu}
\right]}{\left[Q^{2}+yp^{2}\right]^{2}}\sin^{2}\left[\frac{p\times Q}{2}\right]=0.
\end{align}
This is a known result where we see that the NC Chern-Simons
theory is a free theory \cite{das}. However, we conclude that the HD
contributions are sufficient to modify the character of the
polarization tensor of the pure NC Chern-Simons theory, rendering a
nonvanishing result.

Moreover, in order to evaluate the form factors
\eqref{eq:7.2a}--\eqref{eq:7.2d} it is easier to compute the
contraction of the expressions \eqref{eq:8.21a}--\eqref{eq:8.21c}
with the operators $\eta^{\mu\nu}$,
$\tilde{p}^{\mu}\tilde{p}^{\nu}/\tilde{p}^{2}$,
$\varepsilon^{\mu\nu\lambda}p_{\lambda}$, and
$\left(u^{\mu}\tilde{p}^{\nu}+u^{\nu}\tilde{p}^{\mu}\right)$.
Surprisingly, we immediately find out that the following projections
\begin{align}
\left(u^{\mu}\tilde{p}^{\nu}+u^{\nu}\tilde{p}^{\mu}\right)\Pi_{\mu\nu}^{\textrm{(a+b)}}
=\left(u^{\mu}\tilde{p}^{\nu}+u^{\nu}\tilde{p}^{\mu}\right)\Pi_{\mu\nu}^{\textrm{(c)}}&
=\left(u^{\mu}\tilde{p}^{\nu}+u^{\nu}\tilde{p}^{\mu}\right)\Pi_{\mu\nu}^{\textrm{(d)}} =0,
\end{align}
and
\begin{align}
\varepsilon^{\mu\nu\lambda}p_{\lambda}\Pi_{\mu\nu}^{\textrm{(a+b)}}
=\varepsilon^{\mu\nu\lambda}p_{\lambda}\Pi_{\mu\nu}^{\textrm{(c)}}
&
=\varepsilon^{\mu\nu\lambda}p_{\lambda}\Pi_{\mu\nu}^{\textrm{(d)}} =0,
\end{align}
vanish identically. In particular, we have made use of the identities
$p.u=0$ and $p.\tilde{p}=0$ in order to obtain the previous results. So, the
form factors
$\Pi_{\textrm{0}}^{\textrm{\tiny{A}}}$ and
$\Pi_{\textrm{0}}^{\textrm{\tiny{S}}}$,
Eqs.\eqref{eq:7.2c} and \eqref{eq:7.2d}, respectively, vanish
at this order. Hence, we are left to compute only the projection of
\eqref{eq:8.21a}--\eqref{eq:8.21c} onto the operators $\eta^{\mu\nu}$
and $\tilde{p}^{\mu}\tilde{p}^{\nu}/\tilde{p}^{2}$.

The planar and non-planar parts of the contractions $\eta^{\mu\nu}\Pi_{\mu\nu}\left(p\right)$ and $\frac{\tilde{p}^{\mu}\tilde{p}^{\nu}}{\tilde{p}^{2}}\Pi_{\mu\nu}\left(p\right)$ are computed in the Appendix~\ref{appC}.

Before computing the form factors $\Pi_{\emph{\textbf{e}}}$ and
$\widetilde{\Pi}_{\emph{\textbf{e}}}$, Eqs.\eqref{eq:7.2a} and
\eqref{eq:7.2b}, respectively, let us now establish the
renormalizability of the theory. First, note that due to the fact
that the form factors
$\Pi_{\textrm{o}}^{\textrm{A}}$ and
$\Pi_{\textrm{o}}^{\textrm{S}}$ are null,
the expression of the complete propagator \eqref{eq:7.3} reads
\begin{align}
i\mathcal{D}_{\mu\nu} &
=-\frac{\Pi_{\textrm{e}}+\widetilde{\Pi}_{\textrm{e}}}{D}\eta_{\mu\nu}+\left(\frac{\Pi_{\textrm{e}}+\widetilde{\Pi}_{\textrm{e}}}{D}+\frac{\xi}{p^{2}}\right)\frac{p_{\mu}p_{\nu}}{p^{2}}+\frac{\widetilde{\Pi}_{\textrm{e}}}{D}\frac{\tilde{p}_{\mu}\tilde{p}_{\nu}}{\tilde{p}^{2}}+\frac{m\left(1-\frac{p^{2}}{m^{2}}\right)}{D}i\varepsilon_{\mu\nu\lambda}p^{\lambda},\label{eq:8.28}
\end{align}
with a simplified form
$D=\Pi_{\textrm{e}}(\Pi_{\textrm{e}}+\widetilde{\Pi}_{\textrm{e}})-p^{2}m^{2}\left(1-\frac{p^{2}}{m^{2}}\right)^{2}.$
It is sufficient for our interest, in defining the physical pole of
the NC HD Chern-Simons sector, to consider the CP odd term from
\eqref{eq:8.28}
\begin{equation}
i\mathcal{D}_{\mu\nu}\sim\frac{m\left(1-\frac{p^{2}}{m^{2}}\right)}
{p^{2}\left[p^{2}\Pi_{e}(\Pi_{e}+\widetilde{\Pi}_{\textrm{e}})-m^{2}\left(1-\frac{p^{2}}{m^{2}}\right)^{2}
\right]}i\varepsilon_{\mu\nu\lambda}p^{\lambda},\label{eq:8.29}
\end{equation}
where we have performed the scaling $\left\{
\Pi_{\textrm{e}},\widetilde{\Pi}_{\textrm{e}}\right\} \rightarrow p^{2}\left\{\Pi_{\textrm{e}},\widetilde{\Pi}_{\textrm{e}}\right\} $. By a simple analysis and
manipulation we can rewrite the expression \eqref{eq:8.29} in its
renormalized form
\begin{align}
i\mathcal{D}_{\mu\nu} &
\sim\frac{m_{ren}\mathcal{Z}_{_{CP}}}{p^{2}\left[p^{2}-m_{ren}^{2}\right]}i\varepsilon_{\mu\nu\lambda}p^{\lambda},\label{eq:8.30}
\end{align}
where the physical pole $p^{2}=m_{ren}^{2}$ is localized so that the
renormalized mass of the Chern-Simons gauge field and the respective
renormalization constant were defined as
\begin{equation}
m_{ren}=\mathcal{Z}_{_{CP}}m,\quad\mathcal{Z}_{_{CP}}=\frac{1}{\sqrt{1+\Pi_{\textrm{e}}\left(\Pi_{\textrm{e}}+\widetilde{\Pi}_{\textrm{e}}\right)}}.\label{eq:8.31}
\end{equation}

In order to finally compute the renormalized mass, it is worth to
evaluate the form factors $\Pi_{\emph{\textbf{e}}}$ and
$\widetilde{\Pi}_{\emph{\textbf{e}}}$ in the highly
noncommutative limit, i.e. $\beta=p^{2}/m^{2}\rightarrow0$ while
$\tilde{k}^{2}$ is kept finite. This allows us to obtain the leading
noncommutative effects onto the renormalized mass.

Hence, from the expression \eqref{eq:7.2a} we find that
\begin{align}
\Pi_{\textrm{e}}\left(p\right) & =\eta_{\mu\nu}\Pi^{\mu\nu}(p)-\frac{\tilde{p}_{\mu}\tilde{p}_{\nu}}{\tilde{p}^{2}}
\Pi^{\mu\nu}(p)\nonumber\\
 & =\frac{e^{2}}{12\pi}\frac{1}{m^{2}|\tilde{p}|^{3}}
 \biggl\{3\left(2+m|\tilde{p}|\right)^{2}e^{-m|\tilde{p}|}
 -\left(12+m^{3}|\tilde{p}|^{3}\right)\biggr\}.\label{eq:8.32}
\end{align}
Moreover, due to the structure of the renormalization constant
$\mathcal{Z}_{_{CP}}$ in \eqref{eq:8.31}, it shows convenient to
compute the following combination
\begin{align}
\Pi_{\textrm{e}}+\widetilde{\Pi}_{\textrm{e}} & =\frac{\tilde{p}_{\mu}\tilde{p}_{\nu}}{\tilde{p}^{2}}\Pi^{\mu\nu}(p)\nonumber\\
 & =-\frac{e^{2}}{24\pi}\frac{1}{m^{2}|\tilde{p}|^{3}}\biggl\{-24+m^{3}|\tilde{p}|^{3}
 +3\left(8+8m|\tilde{p}|+8m^{2}\tilde{p}^{2}+5m^{3}|\tilde{p}|^{3}\right)
 e^{-m|\tilde{p}|}\biggr\}.\label{eq:8.33}
\end{align}

We can finally make use of the results \eqref{eq:8.32} and
\eqref{eq:8.33} into the expression \eqref{eq:8.31} in order to
compute the renormalized mass
\begin{align}
m_{ren}   \simeq m-\frac{1}{16\pi^{2}}\frac{1}{\kappa^{2}m|\tilde{p}|^{2}}+\frac{5}{32\pi^{2}}
 \frac{1}{\kappa^{2}|\tilde{p}|}-\frac{1}{8\pi^{2}}\frac{m}{\kappa^{2}}+
 \mathcal{O}\left(\frac{m|\tilde{p}|}{\kappa}\right),\label{eq:8.34}
\end{align}
when writing this expression we have used the tree-level relation
$e^{2}\sim m/\kappa$. Moreover, we simplified the above terms by taking the leading noncommutative perturbations $m|\tilde{p}|\ll 1$.
Furthermore, recalling that at the on-shell limit we have
$|\tilde{p}|\rightarrow m\left|\theta\right|$, the
dispersion relation for the Chern-Simons gauge field reads
\begin{align}
\omega^{2} &
\simeq\left|\vec{p}\right|^{2}+m^{2}-\frac{1}{4\pi^{2}}\frac{m^{2}}{\kappa^{2}}
-\frac{1}{8\pi^{2}}\frac{1}{\kappa^{2}m^{2}}\frac{1}{\theta^{2}}
+\frac{5}{16\pi^{2}}\frac{1}{\kappa^{2}}\frac{1}{\theta}
+\mathcal{O}\left(\frac{m^{2}\theta}{\kappa}\right).\label{eq:8.35}
\end{align}
In particular, we see that a correction is obtained for the massive
mode,
$m_{eff}^{2}=m^{2}\left[1-\frac{1}{4\pi^{2}}\frac{1}{\kappa^{2}}\right]$,
and that a leading $1/\theta^2$ and subleading $1/\theta$ UV/IR mixing are present in this
case.

%%%%%%%%%%%%%%%%%%%%%%%%%%%%%%%%%%%%%%%%%%%%%%%%%%%%%%%%%%%%%%%
%%%%%%%%%%%%%%%%%%%%%%%%%%%%%%%%%%%%%%%%%%%%%%%%%%%%%%%%%%%%%%%
\section{Concluding remarks}
\label{sec6}
%%%%%%%%%%%%%%%%%%%%%%%%%%%%%%%%%%%%%%%%%%%%%%%%%%%%%%%%%%%%%%%
%%%%%%%%%%%%%%%%%%%%%%%%%%%%%%%%%%%%%%%%%%%%%%%%%%%%%%%%%%%%%%%

In this paper we have considered the Yukawa field theory for
neutral and charged scalar fields in the framework of a \NC three-dimensional spacetime.
In order to study the dynamics of the scalar fields we have followed
the effective action approach by integrating out the fermionic
fields. The two-point function renormalization for both cases were carefully
established, allowing us to define the physical dispersion
relation and hence study the UV/IR anomaly.

Initially we established the main properties of the effective action
approach for \NC field theories \cite{ref11,bufalo}. In particular,
we defined the \NC Yukawa action for a neutral scalar field and computed
its effective action, and showed that besides its kinetic terms, only a cubic interaction $\phi^3_{\star}$ is present at the leading order ${\cal{O}}(m^{0})$, and that
$\phi^n_{\star}$ (for $n>3$) interacting terms are absent in its
effective action when defined in a three-dimensional spacetime. Additionally, we considered the case of charged
scalar fields, where a gauge field is need in order to ensure a
local gauge invariance. Hence, by computing the respective cubic and
quartic terms, we found the minimal interacting parts of the NC Higgs model,
i.e. charged scalar fields minimally coupled to a gauge field.

In order to illustrate the behavior of the neutral scalar field in
a \NC three-dimensional spacetime, we considered the effective action  with an additional derivative coupling supplementing the cubic interacting term. We then determined the respective Feynman rules
and the one-loop correction to the self-energy, also the
renormalized mass was established. After computing the planar and
non-planar contributions to the one-loop self-energy we found the
physical dispersion relation, where we have showed a correction to the
bare mass and the presence of a $1/\theta$ singular term, as a result of UV/IR instability caused
by \NC effects.

By completeness, we considered the effective action for charged
scalar fields coupled with a dynamical Chern-Simons gauge field. We have considered a higher-derivative contribution to the kinetic term of the gauge field in order to discuss some novel features in the gauge field sector. By computing
first the one-loop correction to the scalar field self-energy, we
showed that this contribution is non-physical due to its dependence on the gauge
parameter $\xi$ onto the dispersion relation; by means of
illustration we presented how the dispersion relation is modified
and became physical if a Maxwell-Chern-Simons propagator is considered instead of the HD-Chern-Simons propagator. Afterwards, we
established the renormalization and analysed the one-loop dispersion
relation for the gauge field. In the resulting expression we found
the presence of a correction to the bare mass parameter, and that a
leading $1/\theta^2$ and subleading $1/\theta$ UV/IR mixing are present in this case.

It is worth of mention that all the previous analysis were
considered in the highly \NC limit of the Chern-Simons-Higgs theory,
that also corresponds to the low-momenta limit. As we have
discussed, this kind of effective models are nowadays of major
interest for physical application in planar materials, in particular
in the description of new materials in the framework of condensed
matter physics, which allows the use of effective low-energy models.
Moreover, this model can be an appropriate field-theoretical framework for study of Aharonov-Bohm effect.
This framework also works in a way to provide a consistent scenario
to scrutinize theoretical features in order to set stringent bounds on deviations of known field theory properties, for instance standard model symmetries.

\subsection*{Acknowledgements}
We would like to thank M.M. Sheikh-Jabbari for discussion. R.B. thankfully acknowledges CAPES/PNPD for partial support, Project No.
23038007041201166.

%%%%%%%%%%%%%%%%%%%%%%%%%%%%%%%%%%%%%%%%%%%%%%%%%%%%%%%%%%%%%%%%%%%%%%%%%%%%%%%%%%%%%%%%%%%%%%%%%%%%%%%%%%%%%%%%%%%%%%%%%%%%%%%%%%%%%%%%%%

\appendix

\section{Non-planar integrals}

\label{appA}

Along the paper we have made use of some results involving momentum
integration. We shall recall some of these results, in particular
those involving a non-planar factor. The simplest non-planar
integration reads
\begin{equation}
\int\frac{d^{\omega}q}{\left(2\pi\right)^{\omega}}\frac{1}{\left(q^{2}-s^{2}\right)^{a}}e^{ik\wedge
q}=\frac{2i\left(-\right)^{a}}{\left(4\pi\right)^{\frac{\omega}{2}}}
\frac{1}{\Gamma\left(a\right)}\frac{1}{\left(s^{2}\right)^{a-\frac{\omega}{2}}}
\left(\frac{|\tilde{k}|s}{2}\right)^{a-\frac{\omega}{2}}
K_{a-\frac{\omega}{2}}\left(|\tilde{k}|s\right).\label{eq:C.1}
\end{equation}
Next, we have the tensor integration
\begin{align}
\int\frac{d^{\omega}q}{\left(2\pi\right)^{\omega}}\frac{q^{\mu}q^{\nu}}{\left(q^{2}-s^{2}\right)^{a}}e^{ik\wedge
q} &
=\eta^{\mu\nu}F_{a}+\frac{\tilde{k}^{\mu}\tilde{k}^{\nu}}{\tilde{k}^{2}}G_{a},\label{eq:C.2}
\end{align}
where we have introduced the following quantities
\begin{align}
\left\{ F_{a},G_{a}\right\}  &
=\frac{i\left(-\right)^{a-1}}{\left(4\pi\right)^{\frac{\omega}{2}}}\frac{1}{\Gamma\left(a\right)}\frac{1}{\left(s^{2}\right)^{a-1-\frac{\omega}{2}}}~\left\{
f_{a},g_{a}\right\} ,\label{eq:C.3}
\end{align}
with
\begin{align}
f_{a} & =\left(\frac{s|\tilde{k}|}{2}\right)^{a-1-\frac{\omega}{2}}K_{a-1-\frac{\omega}{2}}\left(|\tilde{k}|s\right),\label{eq:C.3a}\\
g_{a} &
=\left(2a-2-\omega\right)\left(\frac{s|\tilde{k}|}{2}\right)^{a-1-\frac{\omega}{2}}K_{a-1-\frac{\omega}{2}}\left(|\tilde{k}|s\right)-2\left(\frac{s|\tilde{k}|}{2}\right)^{a-\frac{\omega}{2}}K_{a-\frac{\omega}{2}}\left(|\tilde{k}|s\right).\label{eq:C.3b}
\end{align}

%%%%%%%%%%%%%%%%%%%%%%%%%%%%%%%%%%%%%%%%%%%%%%%%%%%%%%%%%%%%%%%%%%%%%%%%%%%%%%%%%%%%%%%%%%%%%%%%%%%%%%%%%%%%%%%%%%%%%%%%%%%%%%%%

\section{Effective action}

\label{appB}
We summarize in this appendix some useful and important long expressions related to the computation of the effective action in the Sec.~\ref{sec3}.

\subsection{$\phi\phi A$ contribution}
The three distinct contributions for the one-loop effective action for the $\phi\phi A$ vertex \eqref{eq:2.4} are given by
\begin{align}
\Gamma_{\left(i\right)}^{\mu}\left(x_{1},x_{2},x_{3}\right) & =\frac{g^{2}e}{3}\int\frac{d^{3}p_{1}}{\left(2\pi\right)^{3}}
\frac{d^{3}p_{2}}{\left(2\pi\right)^{3}}\frac{d^{3}p_{3}}{\left(2\pi\right)^{3}}\left[\left(2\pi\right)^{3}\delta\left(p_{1}+p_{2}+p_{3}\right)\right]
\exp\left[-i\left(p_{1}.x_{1}+p_{2}.x_{2}+p_{3}.x_{3}\right)\right]\nonumber \\
 & \times\exp\left[-\frac{i}{2}\left(p_{1}\times p_{2}+p_{1}\times p_{3}+p_{2}\times p_{3}\right)\right]\Xi_{\left(i\right)}^{\mu}\left(p_{1},p_{2}\right),
\end{align}
in which $p\times q=\theta^{\mu\nu}p_{\mu}q_{\nu}$, where the explicit expressions $\Xi_{\left(i\right)}^{\mu\nu}$ are readily obtained from the graph Fig.~\ref{oneloopdiagrams3}.

Due to the structure of the momentum integral, and our interest in
finding the leading contribution for the interaction terms, we see
that those correspond to the overall
$\mathcal{O}\left(m^{-1}\right)$ terms, that are precisely the
$\mathcal{O}\left(m^{2}\right)$ and $\mathcal{O}\left(m^{0}\right)$
terms in the numerator of these expressions, that will result in a
(linear) derivative coupling. Hence, from the denominator of the
above expressions we find that
\begin{align}
tr\left[...\right]^{a} & \simeq2m^{2}\left(p-k-3s\right)^{\mu}+\frac{2}{\omega}q^{2}\left(3p-3k-5s\right)^{\mu},\\
tr\left[...\right]^{b} & \simeq2m^{2}\left(p-k-3s\right)^{\mu}-\frac{2}{\omega}q^{2}\left(3k+5s+p\right)^{\mu},\\
tr\left[...\right]^{c} &
\simeq2m^{2}\left(p-k-3s\right)^{\mu}+\frac{2}{\omega}q^{2}\left(k+3p-5s\right)^{\mu},
\end{align}
where we have used the identity
$q^{\mu}q^{\nu}\rightarrow\frac{1}{\omega}q^{2}\eta^{\mu\nu}$.
Therefore, with such considerations we can compute the momentum integration straightforwardly,
and by realizing that $\Gamma\left(2-\frac{\omega}{2}\right)$ and
$\Gamma\left(3-\frac{\omega}{2}\right)$ have no poles when
$\omega\rightarrow3^{+}$, we obtain
\begin{align}
\Xi_{\left(a\right)}^{\mu}\left(p,k\right) & =\frac{i}{8\pi}\int d\xi\left[\frac{\left(3p-3k-5s\right)^{\mu}}{\left(m^{2}-A^{2}\left(p,k\right)\right)^{\frac{1}{2}}}-m^{2}\frac{\left(p-k-3s\right)^{\mu}}{\left(m^{2}-A^{2}\left(p,k\right)\right)^{\frac{3}{2}}}\right],\\
\Xi_{\left(b\right)}^{\mu}\left(p,k\right) & =\frac{i}{8\pi}\int d\xi\left[-\frac{\left(3k+5s+p\right)^{\mu}}{\left(m^{2}-A^{2}\left(p,k\right)\right)^{\frac{1}{2}}}-m^{2}\frac{\left(p-k-3s\right)^{\mu}}{\left(m^{2}-A^{2}\left(p,k\right)\right)^{\frac{3}{2}}}\right],\\
\Xi_{\left(c\right)}^{\mu}\left(p,k\right) & =\frac{i}{8\pi}\int
d\xi\left[\frac{\left(k+3p-5s\right)^{\mu}}{\left(m^{2}-A^{2}\left(p,k\right)\right)^{\frac{1}{2}}}-m^{2}\frac{\left(p-k-3s\right)^{\mu}}{\left(m^{2}-A^{2}\left(p,k\right)\right)^{\frac{3}{2}}}\right].
\end{align}
Finally, we can now determine the leading contributions by taking the
long wavelength limit, i.e. the approximation $m^{2}\gg
A^{2}\left(p,k\right)$ in the above expressions. Hence, proceeding
with this calculation and computing the remaining integration, we find
\begin{align}
\Xi_{\left(a\right)}^{\mu}\left(p,k\right) & \simeq\frac{i}{12\pi}\frac{1}{\left|m\right|}\left(p-k\right)^{\mu},\quad
\Xi_{\left(b\right)}^{\mu}\left(p,k\right)  \simeq-\frac{i}{12\pi}\frac{1}{\left|m\right|}\left(k+2p\right)^{\mu},\label{eq:2.6b}\\
\Xi_{\left(c\right)}^{\mu}\left(p,k\right) &
\simeq\frac{i}{12\pi}\frac{1}{\left|m\right|}\left(2k+p\right)^{\mu}.\label{eq:2.6c}
\end{align}

\subsection{$\phi\phi AA$ contribution}
%%%%%%%%%%%%%
The contribution proportional to the structure of the $\phi\phi AA$ vertex \eqref{eq:2.10} is given by six terms which has the general structure
\begin{align}
\Gamma_{\left(i\right)}^{\mu\nu}\left(x_{1},x_{2},x_{3},x_{4}\right) & =-\frac{g^{2}e^2}{4}\int\frac{d^{3}p_{1}}{\left(2\pi\right)^{3}}\cdots\frac{d^{3}p_{4}}
{\left(2\pi\right)^{3}}\left(2\pi\right)^{3}\delta\left(p_{1}+p_{2}+\cdots+p_{4}\right) \\
 & \times\exp\left[-i\left(p_{1}.x_{1}+\cdots+p_{4}.x_{4}\right)\right]\exp\bigg(-\frac{i}{2}
 \sum_{i<j}p_{i}\times p_{j}\bigg)\Xi_{\left(i\right)}^{\mu\nu}\left(p_{1},p_{2},p_{3}\right),\nonumber
\end{align}
where the integral expressions $\Xi_{\left(i\right)}^{\mu\nu}$ are readily obtained from the graph Fig.~\ref{oneloopdiagrams4}.

It should be remarked that due to our interest  in obtaining the last piece of the (minimal) coupling between scalar and gauge fields, we shall now
concentrate in those contributions independent of the external
momenta in order to complete the derivation of the effective action.
In this way, it is easy to show the following equality among the
contributions
\begin{align}
\Xi_{\left(a\right)}^{\mu\nu}\left(p,k,r\right) & =\Xi_{\left(c\right)}^{\mu}\left(p,k,r\right)=\Xi_{\left(d\right)}^{\mu\nu}\left(p,k,r\right)=\Xi_{\left(f\right)}^{\mu\nu}\left(p,k,r\right),\label{eq:2.12a}\\
\Xi_{\left(b\right)}^{\mu\nu}\left(p,k,r\right) &
=\Xi_{\left(e\right)}^{\mu\nu}\left(p,k,r\right),\label{eq:2.12b}
\end{align}
where we find that
\begin{align}
\Xi_{\left(a\right)}^{\mu\nu}\left(p,k,r\right) & \simeq\Gamma\left(4\right)\int d\zeta\int\frac{d^{3}q}{\left(2\pi\right)^{3}}
\frac{tr\left[\left(\displaystyle{\not}q+m\right)\left(\displaystyle{\not}q+m\right)\left(\displaystyle{\not}q+m\right)\gamma^{\mu}
\left(\displaystyle{\not}q+m\right)\gamma^{\nu}\right]}{\left[q^{2}+M^{2}-m^{2}\right]^{4}},\label{eq:2.13a}\\
\Xi_{\left(b\right)}^{\mu\nu}\left(p,k,r\right) &
\simeq\Gamma\left(4\right)\int
d\zeta\int\frac{d^{3}q}{\left(2\pi\right)^{3}}\frac{tr\left[\left(\displaystyle{\not}q+m\right)\left(\displaystyle{\not}q+m\right)
\gamma^{\mu}\left(\displaystyle{\not}q+m\right)\left(\displaystyle{\not}q+m\right)\gamma^{\nu}\right]}{\left[q^{2}+M^{2}-m^{2}\right]^{4}},\label{eq:2.13b}
\end{align}
with the following definition
\begin{align}
\int
d\zeta&=\int_{0}^{1}dx\int_{0}^{1-x}dz\int_{0}^{1-x-z}dw,\nonumber\\
M^{2}\left(p,k,r\right)&=-\left(xp-\left(z+w\right)k-wr\right)^{2}+xp^{2}+zk^{2}+w\left(r+k\right)^{2}.
\end{align}
Now the numerator of both expressions can be computed with help of
the results
$q^{\mu}q^{\nu}\rightarrow\frac{1}{\omega}q^{2}\eta^{\mu\nu}$ and
$q^{\mu}q^{\nu}q^{\sigma}q^{\rho}\rightarrow\frac{1}{\omega\left(\omega+2\right)}\left(q^{2}\right)^{2}\left(\eta^{\mu\nu}\eta^{\sigma\rho}+\eta^{\mu\sigma}\eta^{\nu\rho}+\eta^{\mu\rho}\eta^{\nu\sigma}\right)$.
The resulting expressions are
\begin{align}
tr\left[...\right]^{a} & =-\frac{10}{\omega\left(\omega+2\right)}\left(q^{2}\right)^{2}\eta^{\mu\nu}+12m^{2}\frac{1}{\omega}q^{2}\eta^{\mu\nu}+2m^{4}\eta^{\mu\nu},\\
tr\left[...\right]^{b} &
=\frac{30}{\omega\left(\omega+2\right)}\left(q^{2}\right)^{2}\eta^{\mu\nu}+4m^{2}\frac{1}{\omega}q^{2}\eta^{\mu\nu}+2m^{4}\eta^{\mu\nu}.
\end{align}
Once again we see that the momentum integration is finite, since
$\Gamma\left(2-\frac{\omega}{2}\right)$,
$\Gamma\left(3-\frac{\omega}{2}\right)$ and
$\Gamma\left(4-\frac{\omega}{2}\right)$ have no poles when
$\omega\rightarrow3^{+}$. Hence, the analysis follows as before, and considering the expansion $m^{2}\gg
M^{2}\left(p,k,r\right)$, we find that the leading contributions
from Eqs.\eqref{eq:2.13a} and \eqref{eq:2.13b} at the
$\mathcal{O}\left(m^{-1}\right)$ terms are
\begin{align}
\Xi_{\left(a\right)}^{\mu\nu}\left(p,k,r\right) & \simeq-\frac{i}{12\pi}\frac{1}{\left|m\right|}\eta^{\mu\nu},\quad
\Xi_{\left(b\right)}^{\mu\nu}\left(p,k,r\right)
\simeq\frac{i}{6\pi}\frac{1}{\left|m\right|}\eta^{\mu\nu}.\label{eq:2.14b}
\end{align}

%%%%%%%%%%%%%%%%%%%%%%%%%%%%%%%%%%%%%%%%%%%%%%%%%%%%%%%%%%%%%%%%%%%%%%%%%%%%%%%%%%%%%%%%%%%%%%%%%%%%%%%%%%%%%%%%%%%%%%%%%%%%%%%%

\section{Dispersion relation}
\label{appC}

Therefore, we can compute with no problems the contraction of
\eqref{eq:8.21a}--\eqref{eq:8.21c} to the operators $\eta^{\mu\nu}$,
resulting the total contribution projection
\begin{align}
\eta^{\mu\nu}\Pi_{\mu\nu}\left(p\right) &   =2ie^{2}\mu^{3-\omega}\int\frac{d^{\omega}q}{\left(2\pi\right)^{\omega}}
 \biggl\{-m^{4}\frac{1}{q^{2}\left(q^{2}-m^{2}\right)\left(\left(q-p\right)^{2}-m^{2}\right)}\nonumber \\
 &-m^{4}\frac{p.\left(q-p\right)}{q^{2}\left(q-p\right)^{2}\left(q^{2}-m^{2}\right)
 \left(\left(q-p\right)^{2}-m^{2}\right)}+\frac{1}{q^{2}}+\frac{p.\left(q-p\right)}
 {q^{2}\left(q-p\right)^{2}}\nonumber \\
 & -\frac{\left(2q-p\right)^{2}}{\left(q^{2}-m^{2}\right)
 \left(\left(q-p\right)^{2}-m^{2}\right)}+2\omega\frac{1}{q^{2}-m^{2}}\biggr\}
 \sin^{2}\left[\frac{p\times q}{2}\right].\label{eq:8.22}
\end{align}
while the projection
$\frac{\tilde{p}^{\mu}\tilde{p}^{\nu}}{\tilde{p}^{2}}\Pi_{\mu\nu}$
yields the contribution
\begin{align}
\frac{\tilde{p}^{\mu}\tilde{p}^{\nu}}{\tilde{p}^{2}}\Pi_{\mu\nu}\left(p\right)
 & =2ie^{2}\mu^{3-\omega}\frac{\tilde{p}^{\mu}\tilde{p}^{\nu}}{\tilde{p}^{2}}\int\frac{d^{\omega}q}{\left(2\pi\right)^{\omega}}\biggl\{-m^{4}\frac{q_{\mu}q_{\nu}}{q^{2}\left(q-p\right)^{2}\left(q^{2}-m^{2}\right)\left(\left(q-p\right)^{2}-m^{2}\right)}\nonumber \\
 & +\frac{1}{q^{2}}\frac{q_{\mu}q_{\nu}}{\left(q-p\right)^{2}}-4\frac{q_{\mu}q_{\nu}}
 {\left(q^{2}-m^{2}\right)\left(\left(q-p\right)^{2}-m^{2}\right)}
 +2\frac{\eta_{\mu\nu}}{q^{2}-m^{2}}\biggr\}
 \sin^{2}\left[\frac{p\times q}{2}\right].\label{eq:8.23}
\end{align}
Let us now compute separately the planar contribution from the non-planar one of the above projections, this is achieved by the
identity $2\sin^{2}[\frac{p\times Q}{2}]=1-\cos \left(p\times Q\right) $. First,
by following the previous procedure to compute the momentum
integration, we find for the planar part of \eqref{eq:8.22} and
\eqref{eq:8.23} the following expressions
\begin{align}
\left(\eta^{\mu\nu}\Pi_{\mu\nu}\right)_{p}\left(p\right) & =-\frac{e^{2}m}{8\pi}\biggl\{\frac{1}{2}\int_{0}^{1}dz\int_{0}^{1-z}dx\frac{1}
{\left(\left(1-x\right)-z\left(1-z\right)\beta\right)^{\frac{3}{2}}}\nonumber \\
 & +\frac{3}{4}\beta\int_{0}^{1}dz\int_{0}^{1-z}dx\int_{0}^{1-z-x}dw\frac{\left(x+z\right)}{\Delta^{5}}
 -\beta\int_{0}^{1}dyy\frac{1}{\sqrt{-y\left(1-y\right)\beta}}\nonumber \\
 & -\int_{0}^{1}dy\left[12\sqrt{1-y\left(1-y\right)\beta}+\left(2y-1\right)^{2}\beta\frac{1}
 {\sqrt{1-y\left(1-y\right)\beta}}\right]+12\biggr\},\label{eq:8.24}
\end{align}
and
\begin{align}
\left(\frac{\tilde{p}^{\mu}\tilde{p}^{\nu}}{\tilde{p}^{2}}\Pi_{\mu\nu}\right)_{p} & =\frac{e^{2}m}{8\pi}\biggl\{-\frac{1}{4}\int_{0}^{1}dz\int_{0}^{1-z}dw\int_{0}^{1-z-w}dx
\frac{1}{\Delta^{3}}\nonumber \\
 & -\int_{0}^{1}dy\sqrt{-y\left(1-y\right)\beta}+4\int_{0}^{1}dy\sqrt{1-y\left(1-y\right)\beta}-4\biggr\},\label{eq:8.25}
\end{align}
where we have introduced the notation
$\Delta^{2}=\left(z+w\right)-\left(x+w\right)\left(1-\left(x+w\right)\right)\beta$,
and again $\beta=p^{2}/m^{2}$. Next, we compute the non-planar
contributions with help of the identities \eqref{eq:C.1} and
\eqref{eq:C.2},
\begin{align}
\left(\eta^{\mu\nu}\Pi_{\mu\nu}\right)_{n-p}\left(p\right) & =\frac{e^{2}m}{16\pi}\biggl\{\int_{0}^{1}dz\int_{0}^{1-z}dx
\frac{e^{-m|\tilde{p}|\sqrt{\left(1-x\right)1-z\left(1-z\right)\beta}}}
{\left(\left(1-x\right)-z\left(1-z\right)\beta\right)^{\frac{3}{2}}}
\left(1+m|\tilde{p}|\sqrt{\left(1-x\right)-z\left(1-z\right)\beta}\right)\nonumber \\
 & +\frac{1}{2}\beta\int_{0}^{1}dz\int_{0}^{1-z}dx\int_{0}^{1-z-x}dw\frac{\left(x+z\right)}{\Delta^{5}}
 \left(3+3\Delta m|\tilde{p}|+\Delta^{2}m^{2}\tilde{p}^{2}\right)e^{-\Delta m|\tilde{p}|}\nonumber \\
 & -2\beta\int_{0}^{1}dyy\frac{e^{-m|\tilde{p}|\sqrt{-y\left(1-y\right)\beta}}}
 {\sqrt{-y\left(1-y\right)\beta}}-8\int_{0}^{1}dy\left[\sqrt{1-y\left(1-y\right)\beta}
 -2\frac{1}{m|\tilde{p}|}\right]e^{-m|\tilde{p}|\sqrt{1-y\left(1-y\right)\beta}}\nonumber \\
 & -2\beta\int_{0}^{1}dy\left(2y-1\right)^{2}\frac{e^{- m|\tilde{p}|\sqrt{1-y\left(1-y\right)\beta}
}}{\sqrt{1-y\left(1-y\right)\beta}}-\frac{4}{m|\tilde{p}|}
 \left(6e^{-m|\tilde{p}|}+1\right)\biggr\},\label{eq:8.26}
\end{align}
and
\begin{align}
\left(\frac{\tilde{p}^{\mu}\tilde{p}^{\nu}}{\tilde{p}^{2}}\Pi_{\mu\nu}\right)_{n-p} & =\frac{e^{2}m}{8\pi}\biggl\{\frac{1}{4}\int_{0}^{1}dz\int_{0}^{1-z}dw\int_{0}^{1-z-w}dx\left[1+\Delta m|\tilde{p}|-\Delta^{2}m^{2}\tilde{p}^{2}\right]\frac{e^{-\Delta m|\tilde{p}|}}{\Delta^{3}}\nonumber \\
 & +\int_{0}^{1}dy\sqrt{-y\left(1-y\right)\beta}~e^{-m|\tilde{p}|\sqrt{-y\left(1-y\right)\beta}}
 \nonumber \\
 & -4\int_{0}^{1}dy\sqrt{1-y\left(1-y\right)\beta}~e^{-m|\tilde{p}|\sqrt{1-y\left(1-y\right)\beta}}
 -\frac{4}{m|\tilde{p}|}e^{-m|\tilde{p}|}\biggr\}.\label{eq:8.27}
\end{align}

%%%%%%%%%%%%%%%%%%%%%%%%%%%%%%%%%%%%%%%%%%%%%%%%%%%%%%%%%%%%%%%%%%%%%%%%%%%%%%%%%%%%%%

\global\long\def\link#1#2{\href{http://eudml.org/#1}{#2}}
 \global\long\def\doi#1#2{\href{http://dx.doi.org/#1}{#2}}
 \global\long\def\arXiv#1#2{\href{http://arxiv.org/abs/#1}{arXiv:#1 [#2]}}
 \global\long\def\arXivOld#1{\href{http://arxiv.org/abs/#1}{arXiv:#1}}

%%%%%%%%%%%%%%%%%%%%%%%%%%%%%%%%%%%%%%%%%%%%%%%%%%%%%%%%%%%%%%%%%%%%%%%%%%%%%%%%%%%%%%%%%%%%%%%%%%%%%%%%%%%

\end{document}